\documentclass[letterpaper,twocolumn,10pt]{article}
\usepackage{usenix-2020-09}

\usepackage{tikz}
\usepackage{amsmath}

\usepackage{microtype}
\usepackage{graphicx}
\usepackage{subfigure}
\usepackage{booktabs} % for professional tables
\usepackage{xspace}
\usepackage{comment}
\usepackage{tikz}
\usepackage[inline]{enumitem}
\usepackage{multirow}
\usepackage{hyperref}

\newcommand*\circled[1]{\tikz[baseline=(char.base)]{
            \node[shape=circle,draw,inner sep=1.2pt] (char) {#1};}}
\newcommand{\name}{FastPersist\xspace}

\begin{document}
%-------------------------------------------------------------------------------

%don't want date printed
%\date{}

\title{\Large \bf \name: Accelerating Model Checkpointing in Deep Learning}
%\title{\Large \bf \name: Accelerating Model Checkpointing in Deep Learning Training via Exploiting Parallelism in NVMe}
%%%%%%%%%%%%%%%%%%%%%%%%%%%%%%%%%%%%%%%%%%%%%%%%%%%%%%%
%for single author (just remove % characters)
\author{
{\rm Guanhua Wang, Olatunji Ruwase, Bing Xie, Yuxiong He}\\
\\
Microsoft DeepSpeed
%Your Institution
%\and
%{\rm Second Name}\\
%Second Institution
% copy the following lines to add more authors
% \and
% {\rm Name}\\
%Name Institution
} % end author
%%%%%%%%%%%%%%%%%%%%%%%%%%%%%%%%%%%%%%%%%%%%%%%%%%%%%%%%%%%%%%%%%%%%%%%%%%
\maketitle

%-------------------------------------------------------------------------------
\begin{abstract}

Model checkpoints are critical Deep Learning (DL) artifacts that enable fault tolerance for training and downstream applications, such as inference. However, writing checkpoints to persistent storage, and other I/O aspects of DL training, are mostly ignored by compute-focused optimization efforts for faster training of rapidly growing models and datasets.  
%However, optimizations efforts for faster training of rapidly growing models and datasets have mainly focused on compute acceleration, thus making IO aspects, such as persisting model checkpoints, a major bottleneck. 
Towards addressing this imbalance, we propose \name{} to accelerate checkpoint creation in DL training. \name{} combines three novel techniques: (i) NVMe optimizations for faster checkpoint writes to SSDs, (ii) efficient write parallelism using the available SSDs in training environments, and (iii) overlapping checkpointing with independent training computations. Our evaluation using real world dense and sparse DL models shows that \name{} creates checkpoints in persistent storage up to 116x faster than baseline, and enables per-iteration checkpointing with negligible overhead.

\end{abstract}

%  Model checkpointing is I/O intensive and a growing training bottleneck as models and compute scale

% Existing approaches make inefficient use of I/O paths to persistent storage

% Propose efficient utilization of I/O paths to accelerate and scale model checkpointing

% Two complimentary techniques:

% i. NVMe optimizations to accelerate single rank checkpoint writing

% ii. Leverage parallel I/O paths (between rank and storage) to scale across multiple ranks/nodes

% Evaluation to demonstrate

% i. 10X speedup for single rank checkpoint writing over baseline torch.save()

% ii. Linear scaling of checkpointing across multiple nodes

% iii. Can checkpoint each model update without impact on TFLOP/GPU (i.e., training efficiency)

\section{Introduction}
%textcolor{blue}{write first two paragraph for larger scope, feel free to reverse back or re-write.} 

In the field of Deep Learning (DL), model checkpoints (i.e., persistent snapshots of model state) play a critical role of providing fault tolerance for training, and enabling downstream applications such as inference, finetuning, and distillation. While DL continues to enable significant progress in a wide range of artificial intelligence domains, such as natural language processing (NLP)~\cite{bert, gpt3, megatron_turing}, image processing~\cite{imagenet, dalle, sd}, and recommendations~\cite{dlrm, tbsm}, these impressive gains require dramatic scaling of model and dataset. For example, state-of-the-art NLP grew from Bert (300M parameters trained on 3B tokens)~\cite{bert} to GPT3 (175B parameters trained on 300B tokens)~\cite{gpt3} in just two years. Such dramatic DL scaling commensurately increases the amount of computation and time required for model training.

\begin{figure}[t]
\centering
\includegraphics[width=0.45\textwidth]{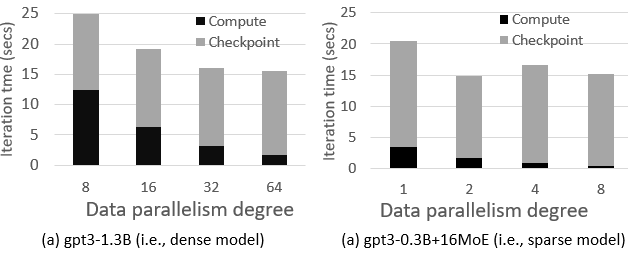}
\caption{\label{fig:dp_checkpoint_overhead}\emph{Impact of data parallelism on training time of (a) dense and (b) sparse models, on up to 128 V100-32GB GPUs.}}
\end{figure}

Optimization efforts to ensure that high-quality DL models can be fully trained in a reasonable amount of time (e.g., few months) have mostly focused on the computation performance of training while ignoring I/O. This imbalanced optimization approach has made I/O, such as data loading~\cite{coordl, nopfs} and model checkpointing~\cite{checkfreq, check-n-run}, a growing bottleneck for large-scale DL training. In Figure~\ref{fig:dp_checkpoint_overhead}, we illustrate this issue for checkpointing using {\it data parallelism} (DP), a popular large-scale training optimization that leverages parallel accelerators to reduce DL computation time. We consider a dense and a sparse GPT3~\cite{gpt3} model, while comparing the computation latency of one training iteration using V100-32GB GPUs ({\it Compute}), versus the latency of checkpointing into locally attached Solid State Device (SSD) storage ({\it Checkpoint}). We observe $\sim 7$X {\it Compute} reduction for both models with DP scaling of $8$ to $64$, and $1$ to $8$ for the dense and sparse models respectively. In contrast, we observe that {\it Checkpoint}, unchanged by DP, increasingly dominates the overall time, growing from $50\%$ to $89\%$, and $82\%$ to $96\%$ for the dense and sparse model respectively. Thus, for large-scale DL, model checkpointing cost can significantly impact end-to-end training time.

In practice, users reduce checkpointing frequency (e.g., every 10s/100s iterations) in order to limit the performance impact on training (e.g., $10\%$ of training time is widely accepted as a reasonably low overhead \cite{xie-ipdps21}). However, this approach also increases the amount of computation and time lost to training interruptions. Thus, the scaling trend of DL beyond 1000s of accelerators~\cite{megatron_turing, bloom176b}, coupled with the high failure rates of large scale systems~\cite{schroeder2009dram,di2014lessons,gupta2017failures} makes low frequency checkpointing strategies less attractive, if still tolerable.

Thus, efficient but frequent checkpointing (e.g., every iteration) becomes desirable for large-scale DL training, and in this work we propose \name{} to provide such a capability. \name{} improves checkpointing performance for DL training by reducing and hiding checkpointing latency. \name{} reduces checkpointing latency by combining two novel and complimentary techniques: (i) optimizations that leverage Non Volatile Memory Express (NVMe) SSDs for faster checkpoint writes from a single rank, and (ii) parallelizing checkpoint writes across the parallel I/O paths between DP ranks and SSDs. \name{} hides checkpointing latency by leveraging domain-knowledge of when the model is updated to overlap checkpoint writes with independent operations (i.e., forward and backward passes) of the next iteration. Since these independent operations typically account for over $90\%$ of compute time, complete overlapping is often possible. 

\iffalse
This work proposes \name{} to improve checkpointing by maximizing the use of hardware I/O bandwidth and leveraging the properties of training parallelism. In particular, \name{} reduces checkpointing latency by combining two novel and complimentary techniques: (i) optimizations that leverage Non Volatile Memory Express (NVMe) SSDs for faster checkpoint writes from a single rank, and (ii) parallelizing checkpoint writes across the parallel I/O paths between DP ranks and SSDs. \name{} avoids checkpoint creation stalls by leveraging the data dependency of DL training to overlap creation with independent operations (i.e., forward and backward passes) of the next iteration. Since these independent operations typically account for over $90\%$ of compute time, complete overlapping is often possible. 
\fi

We evaluate \name{} using micro-benchmarks, as well as dense and sparse GPT3 models on up to $128$ V100-32GB GPUs. The results show that \name{} significantly accelerates checkpoint creation, and is up to $116X$ faster than PyTorch. Also, \name{} efficiently overlaps checkpointing so that checkpointing on each iteration incurs $<~5\%$ overhead. \name{} appears well suited for large-scale DL because it leverages DP to scale checkpointing performance, similar to how DP is used to scale compute performance. Also, \name{} leverages DP to scale I/O hardware utilization and improve efficiency on increasingly larger training hardware clusters. \name{} makes frequent checkpointing a practical solution for minimizing the impact of training interruptions, especially for large-scale DL involving 100s/1000s of servers.  

\iffalse
We evaluate \name{} using micro-benchmarks, as well as dense and sparse GPT3 models on up to $128$ V100-32GB GPUs. The results show that \name{} significantly reduces checkpoint creation latency (e.g., $67$X and $31$X for gpt3-1.3B and gpt3-0.3B+16MoE respectively), and eliminates checkpoint creation stalls for each model even when checkpointing on every iteration. \name{} appears well suited for large-scale DL because it leverages DP to scale checkpointing performance, similar to how DP is used to scale compute performance. Also, \name{} leverages DP to scale I/O hardware utilization and improve efficiency on increasingly larger training hardware clusters. \name{} makes frequent checkpointing a practical solution for minimizing the impact of training interruptions, especially for large-scale DL involving 100s/1000s of servers.  
\fi

The key contributions of this paper are the following:

{\bf 1}. To the best of our knowledge, \name is the first system to optimize DL model checkpointing through efficient utilization of NVMe SSDs. 

{\bf 2}. We design efficient algorithms for writing checkpoint data from accelerator memory into NVMe SSD, and for parallelizing checkpoint writes among DP ranks. 

%at peak device bandwidth. Our solution further includes parallel model checkpointing and achieves near-linear scalability on write throughput.

{\bf 3}. We implement \name{} using popular DL frameworks, PyTorch and DeepSpeed. 

{\bf 4}. We evaluate \name{} with real world DL models and show it incurs negligible overhead to checkpoint every iteration. We plan to open source \name{} to DeepSpeed soon. %\textcolor{blue}{do we really need this bullet point?}.

%We achieve up to 116x speedup over conventional checkpointing solution for giant dense models and MoE models.

% Why is model checkpointing important?

% Model checkpointing is very slow. 

% Our solution - efficiently use available I/O hardware

% Why focus on only on writes, not reads? Solution does not hurt read performance

\section{Background} 
%We provide a brief background of relevant aspects of model training and checkpointing. 

In distributed DL, a typical training run executes an iterative learning method (e.g., SGD, ADAM) among a group of {\it ranks}, where each rank represents a different framework process (e.g., a PyTorch process) and executes on a different GPU \footnote{Theoretically, DL frameworks generally support multi-GPUs per rank. But in production use, most users let each rank use a different GPU.}.  
Ranks process batches of data samples %{\it micro batch size} (data samples) of a {\it global batch size}(training dataset) 
parallely and iteratively. In an iteration, each rank performs a three-step 
 computational procedure:
\begin{enumerate*}
\item a forward pass to compute loss.
\item a backward pass to compute gradients of the model parameters.
\item an optimization step to update parameters.
\end{enumerate*}

\subsection{Checkpointing in Distributed Deep Learning Training}
\label{subsec:background_training}

To tolerate interruptions from failure-prone environment, users checkpoint model  states after the optimization step of each one or more iterations. 
Generally speaking, checkpointing in distributed DL training is new and cannot be addressed easily by the existing approaches proposed for High Performance Computing (HPC) applications \cite{hpc-checkpoint1, hpc-checkpoint2, hpc-checkpoint3}, largely due to the uniqueness of training parallelism (\S\ref{subsubsec:background_parallel}), the use of gradient accumulation  (\S\ref{subsubsec:background_batch}), and the serialization in checkpoint state (\S\ref{subsubsec:background_state}).   

% In the following discussions, we use the acronyms MBS and GBS to refer to micro batch size and global batch size respectively, and use rank and GPU alternatively.

\iffalse
Model training is an iterative process of updating the model parameters from a randomly initialized state by processing data samples of a global batch size (GBS) drawn from a dataset. Each iteration comprises of a {\it forward pass} that uses the model parameters to compute a loss value for an input, a {\it backward pass} that uses the loss to compute gradients, and an {\it optimizer pass}~\cite{sgd,adam,adagrad} that uses the gradients to update the model parameters. The resource-intensive and time-consuming nature of model training makes it prone to interruptions, and so checkpointing to persistent storage is critical for recovery and completion. Since the most intuitive time to creating checkpoints is after model update, checkpointing typically occurs at iteration intervals.  
\fi

\subsubsection{Training Parallelism}
\label{subsubsec:background_parallel}

%With the exponential growth of datasets and models, data parallelism (DP) and model parallelism (MP) are devised  to facilitate and advance ever-larger DL training. As a result, composing these techniques has enabled model scientists to grow datasets from hundreds of GBs to tens of TBs
%and at the same time grow the largest DL models from $300$M parameters in 2018 \cite{bert} to $530$B  parameters in 2022 \cite{megatron_turing} . %However, beyond advancing training efficiency, these techniques also bring checkpointing into a new era. 

To speedup model training process, DP and model parallelism (MP) are widely adopted , where DP replicates model on each rank and let ranks train on different batches in parallel, and MP splits a model across ranks and let different ranks train on the same data jointly.

%Specifically, DP is introduced to address the cases where a GBS is too large to fit into the memory of a single GPU. With DP in use, a GBS is partitioned into MBS chunks, with each one or more chunks assigned to a different rank for parallel processing of model replicas. Clearly, across iterations in a DP-based run, different ranks each maintains the entire model state identically. Thus, when checkpointing in these runs,  it is simply a single rank (usually the first rank) that creates a single checkpoint file periodically.  

%Comparatively, MP is introduced to address the cases where a model is too large to fit into the memory of a single GPU.  In a typical MP-based run, a single model is partitioned into a number of disjoint {\it slices}, with each one or more slices executed by a different rank with a copy of the GBS. Different to the single checkpoint in DP, the ranks in MP contain disjoint slices of the model and model state (e.g., parameters, layers). Thus, in the context of checkpointing, each MP rank generates a separate checkpoint file for a separate slice state the rank executes periodically.        

Recently, users combine DP and MP to support larger datasets and models \cite{bert,megatron_turing,gpt3}. In such runs, a model is first partitioned into $n$ slices as MP does. Next, a slice is replicated and forms {\it slice replicas} as DP does with each DP group associated with a different slice. 
Finally, a global batch is partitioned into multiple micro-batches with each micro-batch replicated and assigned to a different DP group. 
%a GBS is partitioned into a number of MBSes as a DP-based run does, with each one or more MBSes  replicated and assigned to one of the $n$ slice replica groups.  
Clearly, when checkpointing in DP/MP combined runs, one rank out of the replicas of a separate slice (usually the first rank in the slice) generates a separate checkpoint file for the slice periodically.   

% In this study, we focus on the checkpointing in DP. We expect our proposals in \name can be easily extended to manage the MP and MP/DP combined training since they are simply multiplications on generating single checkpoint files. 

\subsubsection{Gradient Accumulation in Large Batch Training}
\label{subsubsec:background_batch}

Clearly, DP and MP work efficiently when GPUs are abundant. Unfortunately, in many cases, GPUs are limited, %sparse resources, and training is processed on limited GPUs,  
leaving large batches failed to fit into the memories of limited GPUs.   
To address this memory limitation, 
gradient accumulation (GA) is introduced to allow GPUs continously process multiple batches and only update model once with accumulate gradients over these batches. 
%let a GPU process a large batch as a sequence of MBS chunks.   Particularly, in an iteration, a GPU performs multiple rounds of forward/backward passes with each round on a different MBS, and next it operates on a single optimizer pass that updates the model with the accumulated gradients generated by the aforementioned rounds.
% Clearly, GA works in between of the two extremes of the GPU availability:
% \begin{enumerate*}
% \item      a single GPU processes all of the MBs of a GBs,
% \item a large number of GPUs each processes a single MBs of a GBs.
% \end{enumerate*}
%  When more GPUs available in the system, a fewer number of MBSes will be processed per GPU per iteration, yielding a shorter computation time of the iteration. 
% In our study, we propose \name to minimize checkpointing overhead for DL training at scale. In particular, 
For a given model executed on a fixed set of GPUs, %given system, 
 the checkpointing state and cost are relatively consistent as they are determined by the fixed model state and hardware bandwidth capacity. With a higher degree of GA, the checkpointing overhead can be less visible, but failure recovery is more expensive. Thus, we mainly focus on training with lower degree of GA.

\subsubsection{Model Checkpoint State}
\label{subsubsec:background_state}

In a distributed DL training, a model checkpoint state, abbreviated as a {\it checkpoint state}, maintains the information to restore a model and its state after interruptions. Particularly,  a checkpoint state comprises of model parameters (weights and biases), optimizer parameters (e.g., momentum), data loading iterator, and learning rate schedules, etc.

Creating a checkpoint state in standard DL frameworks, such as PyTorch~\cite{pytorch_osdi}, typically involves tensor serializations before writing to persistent storage. Particularly, tensor serialization provides multiple benefits, including reduced storage footprint, checkpoint portability, and simplified loading logic. The serialization process augments the checkpoint with tensor metadata, such as data type, data size, and originating device (e.g., GPU rank). Thus, the checkpoint creation for DL training is not a single write operation of the entire checkpoint state but rather a sequence of writes of serialized tensors. 

The checkpoint-state size is largely determined by the parameters of the model and the optimizer in use, and is relatively consistent and predictable. In particular,  
large models with billions of parameters are commonly trained in a mixed-precision fashion, where the model parameters are low-precision values (FP16 as $2$ bytes), while the optimizer parameters are full-precision values (FP32 as $4$ bytes). Moreover, these large models are primarily trained using the ADAM optimizer~\cite{adam}, which maintains $12$ bytes for each model parameter. Thus, the checkpoint size of large models in bytes is roughly $14$X of the parameter count. Given fixed hardware bandwidth capacity, the achievable checkpointing throughout can be easily estimated and utilized to maximally overlap training computations. We return to discussion as \S\ref{sec:pipeline-write}.

\subsection{Flash-based Storage}
\label{subsec:background_storage}
Training large DL models is an extremely compute-intensive workload that in practice requires large HPC clusters of muti-GPU server machines~\cite{jean_zay,nvidia_selene}. To provide high I/O bandwidths to GPU-accelerated workloads, 
such clusters are also commonly equipped with flash-based storage, in the form of NVMe SSDs. These SSDs are typically configured as local attached storage, or remote disaggregated storage\cite{flash_remote, wekafs}.

Previous efforts have been made to exploit the bandwidth capacities of SSDs for various workloads.  
For example, Bergman et.al. \cite{spin} identified the programming and performance challenges for fast data transfer between GPU and SSDs, and proposed SPIN to integrate P2P into the OS file I/O stack for better I/O performance. Another example is RocksDB \cite{rocksdb}, a key-value store carefully designed for SSDs.  
These efforts suggest that, flash-based SSD devices can help improve the end-to-end performance significantly with careful design to address the specific needs of workloads and to fit the SSDs properties.
Similar to these efforts,  
we build \name to leverage the bandwidth capacity of SSDs for  checkpointing in distributed DL training.

\section{Limitations of Existing Approaches}
\iffalse
Our optimization goal for \name{} is to improve model checkpointing efficiency such that checkpointing is cheap enough to occur frequently (e.g., checkpointing per iteration)   
with little or no impact to training speed. %We pursue a two-step strategy to achieve this goal: (i) reduce checkpointing latency, and (ii) overlap checkpointing with independent training operations.  
In this section, we discuss two optimization opportunities for this strategy. 
\fi
\begin{figure}[t]
\centering
\includegraphics[width=0.35\textwidth]{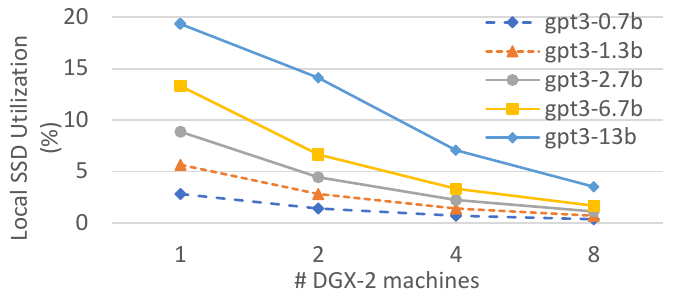}
\caption{\label{fig:torch_ckpt_util}Model checkpointing with \texttt{torch.save()}.}
\end{figure}

In this section, we identify limitations of existing DL checkpointing approaches along three dimensions: (i) SSD write efficiency, (ii) Checkpoint decoupling, and (iii) cost of recovery.   

%This section mainly discusses PyTorch's checkpointing performance and realization.  In particular, PyTorch supports checkpointing with a function namely \texttt{torch.save()} that builds on traditional I/O system libraries with little optimizations toward NVMe devices and utilizes the serialization realized by the Pickle library \cite{meta-pytorch}.  Another popular DL framework, Tensorflow, takes a similar approach that utilizes a callback function, namely \texttt{ModelCheckpoint()} for checkpoint operations  with the use of the same lower-level I/O libraries \cite{google-tensorflow}. Previous studies \cite{2020study} experimented on the checkpointing of these two frameworks in real workloads by varying scaling scenarios. The results show that, PyTorch outperforms Tensorflow in most cases especially when scaling up on models, datasets and GPUs.   

\subsection{Inefficient SSD Writes}
\label{subsec:oppor_performance}

Popular DL frameworks provide standard functions for model checkpointing, such as \texttt{torch.save()} in PyTorch~\cite{meta-pytorch} and \texttt{ModelCheckpoint} in TensorFlow~\cite{google-tensorflow}. These functions build on traditional I/O system libraries with little optimizations for NVMe SSDs. Prior work indicate that \texttt{torch.save} often provides better performance~\cite{2020study}. We measure the  throughput of \texttt{torch.save()} using five GPT3 dense models (Table~\ref{tb:gpt-setup}) and the local SSDs in our cluster(\S\ref{subsubsec:hardware}), and report the results as a percentage of the available SSD peak write throughput (i.e., $24.8$ GB/sec/machine) in Figure~\ref{fig:torch_ckpt_util}. Since these five models are trained with different MP degrees, they present different degrees of write parallelism (\S\ref{subsubsec:background_parallel}).

%For example, {\it gpt3-0.7b} and {\it gpt3-13b} are trained with MP=$0$ or =$1$, and with MP=$16$. Accordingly, 
%in each checkpointing operation,  1 ({\it gpt3-0.7b}) and 16 ({\it gpt3-13b}) parallel instances of \texttt{torch.save()} are used to create $1$ and $16$ files respectively.  
%Figure~\ref{fig:torch_ckpt_util} presents the results.

We first look at {\it gpt3-0.7b} for the write performance of a single GPU on a single machine. We observe that on this setting, only $\sim 3\%$ of the SSD deliverable performance of a node is utilized. Since the $10$GB checkpoint size is large enough to enable efficient I/O, we believe the delivered bandwidth of a single GPU on a single machine is low. %We addresses this limitation in ~\S\ref{subsec:accel_write}. 

Second, we focus on the results of {\it gpt3-13b} for its performance of write parallelism in a machine. Clearly, the {\it gpt3-13b} achieves $\sim 7X$ better performance than the {\it gpt3-0.7b} does in a machine. But considering the $16X$ more parallel writes in {\it gpt3-13b}, it suggests that 
 the individual writes for {\it gpt3-13b} are less efficient than the single write for {\it gpt3-0.b}, indicating the parallelism inefficiencies in a single machine. 
 
Finally, we take a look on the performance of scaling. We observe that the peak performance is $<20\%$ of the hardware bandwidth capacity for all five models.  
Moreover, when scaling up to 8 machines, the SSD write bandwidths are severely underutilized, 
which is concerning given the importance of hardware scaling for optimizing large-scale model training. 

\noindent {\bf Observation \circled{1}: Inefficient SSD Writes}. We observe low bandwidth utilization in the exising checkpointing scheme for individual writes, and parallel writes in a machine and across machines. These inefficienties root from the use of traditional I/O libraries and poor design on write parallelism, motivating us to design new scheme to better exploit NVMe capabilities (\S\ref{subsec:accel_write}) and the scaling characteristics of DL training (\S\ref{subsec:parallel_write}).

\begin{figure}[t]
\centering
\includegraphics[width=0.4\textwidth]{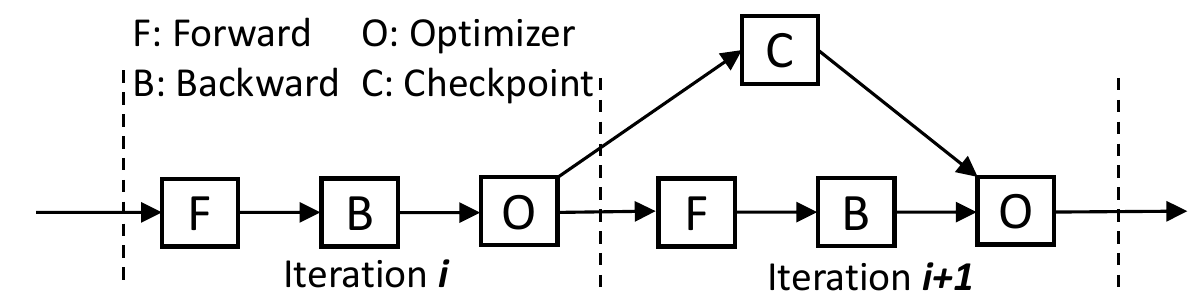}
\caption{\label{fig:train_data_dep}Data dependencies in training iterations.}
\end{figure}

\subsection{Ineffective Checkpoint Decoupling}
\label{subsec:oppor_hide_latency}

\iffalse
Although leveraging the above acceleration opportunities can significantly reduce checkpoint latency, the reality is that checkpointing will still incur a non-zero latency that delays the next training iteration. And so we consider another opportunity for reducing (or eliminating) checkpointing stall, which is to leverage the data dependency of training to decouple checkpointing from independent operations of the subsequent iteration~\cite{checkfreq, check-n-run, veloc}. \fi 

Prior work has leveraged data dependencies of DL training to improve performance by decoupling checkpointing from independent operations in subsequent iterations ~\cite{checkfreq, check-n-run, veloc}. Figure~\ref{fig:train_data_dep} illustrates these dependencies using two iterations ($i$, $i+1$), and arrows to connect data dependent components. We see that checkpointing (C) has no data dependency with either {\it forward} (F) or {\it backward} (B) pass, and so can execute independently of them. In contrast, checkpointing (C) is data dependent on {\it optimizer} (O), because it reads the model updates created by {\it optimizer}, and so both require synchronization. However, we observe two issues with prior work: (i) risk of data loss, and (ii) unsuitability for frequent checkpointing.  

%Another opportunity is to decouple checkpointing from the subsequent iteration by leveraging the data dependencies of training ~\cite{checkfreq, check-n-run, veloc}. Figure~\ref{fig:train_data_dep} illustrates these dependencies using two iterations ($i$, $i+1$), and using arrows to connect data dependent components. To simplify the analysis, we assume checkpointing on each iteration. As is shown clearly, checkpointing (C) has no direct dependency with either the {\it forward} (F) or {\it backward} (B) pass, and accordingly it is safe to execute {\it forward} or {\it backward} independently. In contrast, checkpointing is data dependent on {\it optimizer} (O), % because {\it optimizer} updates the model while checkpointing reads the updates. And 
%thus, it is necessary to serialize {\it optimizer} and checkpointing appropriately. %are properly synchronized.

%Decoupling is previously proposed for checkpointing~\cite{checkfreq, check-n-run, veloc}, but \name{} realizes it differently.  

First, prior work~\cite{checkfreq} splits checkpointing into two phases: (i) a {\it snapshot} phase that writes checkpoints into volatile memory (e.g., accelerator or CPU memory), and (ii) a {\it persist} phase that writes checkpoints into persistent storage. While the {\it snapshot} phase is synchronized with the {\it optimizer} pass, the {\it persist} phase is completely asynchronous. In contrast, \name{} writes checkpoints  directly to persistent storage, and is synchronized with {\it optimizer} pass. Thus, while prior work incurs the risk of losing checkpoint {\it snapshots} to training interruptions, no such risk exists with \name{}. 

Second, prior work~\cite{checkfreq,check-n-run} targets infrequent checkpointing, e.g., intervals of tens or hundreds of iterations. In contrast, \name{} is designed for the more challenging case of frequent checkpointing (e.g., every iteration). We conduct a simple analysis to estimate the minimum checkpoint write bandwidth required to overlap checkpointing for an arbitrary model, and to show that the required bandwidth can be met using a reasonable number of SSDs. 
We express the bandwidth objective for a given model {\it M} in equation~\ref{eq:target-ckpt-bandwidth}, where $T_{F}$, and $T_{B}$ represent the latency of {\it forward}, and {\it backward} respectively, $S_{C}$ represents the checkpoint size, and $B_{C}$ represents the target write bandwidth. This means that if the checkpoint can be created with $B_{C}$ then this creation will overlap the {\it forward} and {\it backward} of the next iteration. 

\begin{equation}
\label{eq:target-ckpt-bandwidth}
    B_{C}(M) >= \frac{S_{C}(M)}{T_{F}(M) + T_{B}(M)}
\end{equation}

Note that equation~\ref{eq:target-ckpt-bandwidth} is constrained in the sense that it assumes a specific distributed training configuration where GBS, DP, GA, and hardware are pre-determined. Thus, for a given training configuration, $T_{F}$ and $T_{B}$ can be empirically obtained through measurements of a few training iterations. For a given model, $S_{C}$ is fixed and independent of distributed training configuration, and could be measured once or estimated using the model parameter count and optimizer type.   

We obtain $B_{C}$ estimates for a training configuration with the most stringent checkpointing latency, and thus the most demanding bandwidth requirements. For each model, this means using the largest valid DP for the published GBS (Table~\ref{tb:gpt-setup}), which minimizes iteration time. The required number of DGX-2 nodes for these configurations are reported in Table~\ref{tb:dense-ckpt-bandwidth}. Since we have only $8$ DGX-2 nodes, fewer than required, we estimate $T_{F}$ and $T_{B}$ using training iterations without GA. This gives smaller $T_{F}$ and $T_{B}$ values compared to the required number of nodes due to cheaper gradient reduction. 
We report the $B_{C}$ estimates in Table~\ref{tb:dense-ckpt-bandwidth}. We observe that although $B_{C}$ increases with model size, the available SSD bandwidth also increases since node count also increases with model size. Moreover, the available SSD write bandwidth on the required node count is larger than $B_{C}$ (maximum of hundreds of GB/sec).

\noindent {\bf Observation \circled{2}: Ineffective Decoupling}. Prior work risks data loss by using volatile memory, and is unsuitable for frequent checkpointing. However, with efficient SSD utilization and decoupling, checkpointing latency can be hidden, even on every iteration, without sacrificing data safety. 

\begin{table}
\small
 \caption{Estimated required write bandwidth ($B_{C}$) to hide checkpoint latency for maximum DP with DGX-2 machines.}
\begin{center}
\label{tb:dense-ckpt-bandwidth}
\begin{tabular}{c|c|c|c} 
 \hline
  Model & DP & \# Nodes & $B_{C}$ (GB/sec) \\
\hline
gpt3-0.7B  & 256 & 16 & 34 \\
\hline
gpt3-1.3B  & 512 & 64 & 59 \\ 
\hline
gpt3-2.7B & 512 & 128 & 81 \\ 
\hline
gpt3-6.7B & 1024 & 512 & 160 \\ 
\hline
gpt3-13B & 1024 & 1024 &  28 \\ 
 \hline
\end{tabular}
\end{center}
\end{table}

\subsection{High Recovery Costs}
\label{subsec:recovery-overhead}
After an interruption, a DL training job resumes by loading the most recent checkpoint. Because of infrequent checkpointing, the checkpoint could have been created tens or hundreds of iterations prior to the interruption, which means those iterations must be repeated as part of recovery. Assuming DL training with $m$ GPUs, checkpointing every $n$ iterations, and iteration time of $t$. Assuming training could interrupted at any point in the $n$ iterations with uniform probability, then the average job recovery overhead is estimated by Equation~\ref{eq:recovery}. 

%When job is interrupted, we resume it by loading the latest version of checkpoint. We assume to use $m$ GPUs together. The model checkpointing happens every $n$ iterations, where each training iteration takes $t$ time. Assuming job failure may happen among these $n$ iteration with uniformed probability, the average job recovery overhead is 

\begin{equation}
\label{eq:recovery}
    \frac{n}{2} * m * t
\end{equation}

For large-scale DL training with infrequent checkpointing (where $n$ is 10s/100s iterations), Equation~\ref{eq:recovery} represents a significant overhead since $m$ could be $1000s$ of GPUs, and $t$ could be 10s of seconds~\cite{megatron_turing, bloom176b}. Since $m$ and $t$ are typically fixed for a training job, the only way to reduce this overhead is to decrease $n$ (i.e., increase checkpointing frequency). In particular, the overhead is minimized when $n=1$, i.e., checkpointing on every iteration.  This is a key motivation of enabling frequent checkpointing for large-scale DL.

\noindent {\bf Observation \circled{3}: Expensive recovery}. The infrequent checkpointing strategy of prior work (i.e., 10s/100s of iterations) incurs high recovery costs for large-scale DL training.

\section{\name Design}
\label{sec:design}

\begin{figure}[h]
\centering
\includegraphics[width=0.49\textwidth]{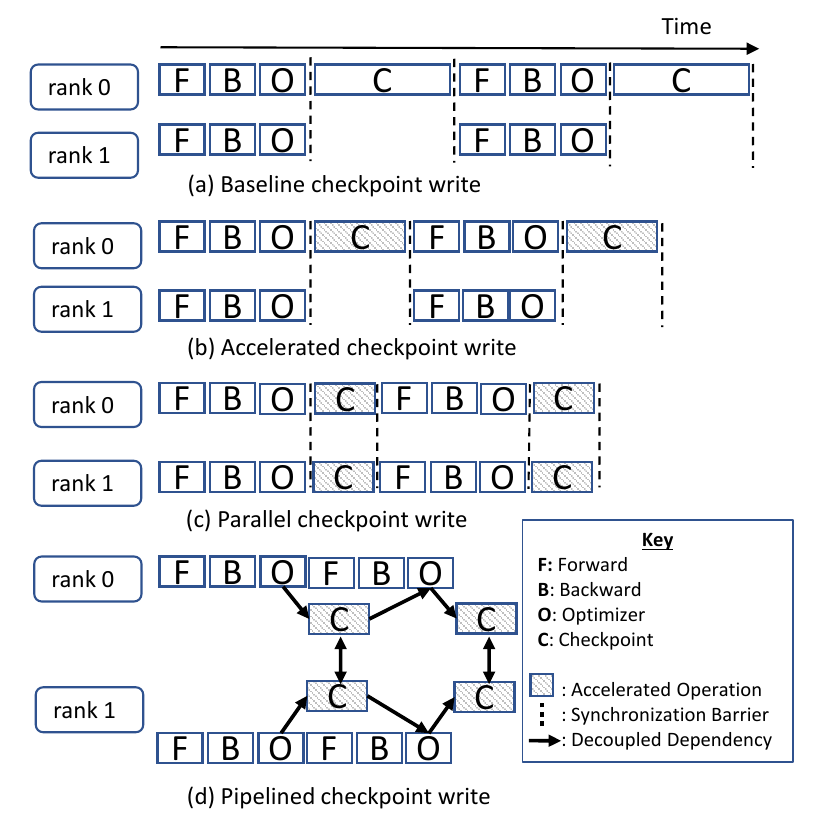}
\caption{\label{fig:system_overview}\emph{Comparing (a) baseline checkpointing against \name{} using two training iterations and DP=2. \name{} improves checkpointing efficiency with three techniques: (b) NVMe-based acceleration, (c) parallelism, and (d) pipelining.}} 
\end{figure}

\name{} combines three optimizations to improve checkpointing of DL training in NVMe devices: (i) acceleration of checkpoint writes, (ii) parallelizing checkpoint writes over DP ranks, and (iii) pipelining checkpoint writes. % with independent operations of the next training iteration. 
Figure~\ref{fig:system_overview} presents a high-level illustration of these optimizations compared to baseline checkpointing. The setting for this comparison is two training iterations of DP training on two accelerators, where checkpoint is created for each iteration. For the baseline case (i.e., Figure~\ref{fig:system_overview}(a)), {\it rank0} creates the checkpoint at the end of the first iteration, while {\it rank1} stalls so that both ranks commence the next iteration simultaneously. % We will now incrementally describe the three \name{} checkpoint creation optimizations.  

% \begin{figure*}[h]
% \centering
% \includegraphics[width=\textwidth]{figures/system_overview.png}
% \caption{\label{fig:system_overview}\emph{Comparing (a) baseline checkpointing against \name{} using two training iterations and DP=2. \name{} improves checkpoint write efficiency with three techniques: (b) NVMe-based acceleration, (c) parallelism, and (d) pipelining.}} 
% \end{figure*}

\subsection{Accelerating Checkpoint Writes}
\label{subsec:accel_write}
The first optimization leverages the multi-gigabyte write bandwidths of NVMe devices. % to reduce the latency of writing checkpoint data from accelerator memory to persistent storage. 
 As shown in Figure~\ref{fig:system_overview}(b), this optimization reduces checkpoint stall and overall iteration time compared to baseline. It is inspired by the observation that traditional I/O system libraries, used by existing DL frameworks are not designed to exploit the performance capability of NVMe devices. Differently, \name{} relies on newer I/O libraries (e.g., {\it libaio}~\cite{libaio} and {\it io\_uring}~\cite{io_uring} in Linux) that are designed with asynchronous and parallelism optimizations for extracting maximum NVMe performance. Obtaining these performance gains, however, requires dealing with more complex programming abstractions than traditional I/O libraries.
In particular, \name{} addresses  restrictions relating to memory buffers and data sizes in the following manner.

\iffalse
As shown in Figure~\ref{fig:system_overview}(b), this optimization reduces checkpoint stall and overall iteration time compared to baseline. This optimization is inspired by the observation that traditional I/O system libraries, which are used by existing DL frameworks, are not designed to exploit the full performance capability of NVMe devices. Thus, \name{} relies on newer I/O libraries (e.g., {\it libaio}~\cite{libaio} and {\it io\_uring}~\cite{io_uring} in Linux) that are designed with asynchronous and parallelism optimizations for extracting maximum NVMe performance. Obtaining these performance gains, however, requires dealing with more complex programming abstractions than traditional I/O libraries. In particular, \name{} addresses  restrictions relating to memory buffers and data sizes in the following manner.
\fi

\begin{figure}[ht]
\centering
\includegraphics[width=0.27\textwidth]{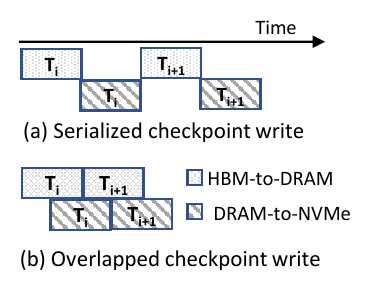}
\caption{\label{fig:double_buffer}\emph{Writing tensors $T_i$ and $T_{i+1}$ from accelerator memory to NVMe. Instead of (a) serializing the two data transfers, (b) \name overlaps them by double-buffering the DRAM.}}
\end{figure}

\textbf{Memory buffer restrictions:} The input data must reside in a memory buffer that allows direct memory access (DMA)  transfer to NVMe memory. In commodity systems, such buffers are page-locked CPU memory because DMA between accelerator and NVMe is not yet broadly available despite the recent significant advances~\cite{spin2017,nvidia_gds}. Consequently, there are two data transfers involved in checkpoint writes: (i) from accelerator memory to page-locked CPU memory, and (ii) from page-locked CPU memory to NVMe memory. Serializing these two data transfers, as illustrated in Figure~\ref{fig:double_buffer}(a), limits write performance. Thus, \name{} employs double buffering of the page-locked CPU memory to overlap the data transfers and improve write performance, as shown in Figure~\ref{fig:double_buffer}(b). In particular, \name{} initiates the second data transfer once the first transfer is halfway done, which essentially eliminates the extra transfer latency. %\textcolor{blue}{If we have more space, maybe elaborate a little more on double buffer with Figure}

\textbf{Data size restrictions:} The input data size is required to be of a certain alignment (e.g., 512-byte boundary in Linux), due to the use of DMA, and the block size of SSDs. This could hinder checkpointing  in two ways. First, checkpoints which do not meet this requirement are excluded from the optimization. Second, checkpoint creation typically involves multiple disk writes of serialized tensors (\S~\ref{subsubsec:background_state}), some of which might fail the alignment requirement. Although padding could be used to meet alignment requirements in both cases, it is undesirable because of checkpoint bloat, as well as breaking compatibility and complicating checkpoint loading. Instead, \name{} addresses this issue in a different way. 

For the first case, \name{} supports checkpoints of arbitrary size by conceptually partitioning the checkpoint data into two: (i) a prefix made up of the largest data subset that meets the alignment requirements, and (ii) a suffix that fails the alignment requirement. This partitioning scheme results in the suffix making up an insignificant portion of the GB-sized checkpoint data for large models (e.g., $<512$ bytes in Linux). \name{} writes the checkpoint prefix using NVMe-optimized libraries, and the suffix using traditional I/O libraries, into the same checkpoint file (preserving compatibility).  This approach achieves virtually all the NVMe performance benefits because of the negligible suffix size.  

For the second case, \name{} aggregates the serialized tensors into a queue of pending bytes that is routinely flushed when alignment requirement is satisfied. This approach could result in the bytes of a tensor being persisted by different write operations, and the bytes of different tensors being persisted by the same write operation. Nevertheless, \name{} preserves correctness by ensuring that the order in which tensors (and their bytes) are persisted to disk remains unchanged. 

\subsection{Parallelizing Checkpoint Writes}
\label{subsec:parallel_write}
The second optimization leverages the parallel I/O hardware of DP ranks, such as the local SSDs across machines, to further reduce checkpoint write latency. As shown in Figure~\ref{fig:system_overview}(c), this optimization improves performance with two key changes. First, it divides checkpoint creation  among DP ranks so that each rank writes only a portion of checkpoint data, as opposed to {\it rank 0} writing the entire checkpoint. Second, it uses more I/O hardware for checkpoint creation as opposed to using only the I/O hardware of {\it rank 0}. As a result of these changes, checkpointing performance and efficiency is improved in two ways: (i) reduced latency through the increased write bandwidth from using more I/O hardware simultaneously for checkpoint creation, and (ii) increased utilization of I/O hardware. 

This optimization is inspired by the observation that DP ranks (a.k.a., model replicas) hold identical checkpoint data, and so each rank can create any subset of the checkpoint data. This means that it is relatively easy to ensure that the entire checkpoint data is persisted in storage for any partitioning scheme, which is a correctness requirement. However, achieving high efficiency in terms of reducing latency proportionally to I/O hardware scaling is more challenging and requires careful handling of communication, load balancing, and hardware efficiency. We now describe how \name{} addresses these issues in order to realize highly efficient write parallelism. 

\textbf{Communication:} A potential source of overheads during parallel checkpoint writing is communication or coordination among the participating DP ranks. \name{} avoids such overheads by  partitioning the checkpointing task among DP ranks during the distributed model training setup, i.e., before the first training iteration. This means that during checkpoint creation, each DP rank already knows the checkpoint data portion that it is responsible for writing to storage, and can therefore complete its task without communicating with others. This approach assumes that the work partitioning among DP ranks will be fixed for the entire training run. However, a number of events could occur during training that violate this assumption, such as model changes (e.g., parameter freezing) or hardware failures. Nevertheless, we expect \name{} to work well in practice, since such events would generally require a repeat of the training setup, creating the opportunity to recompute the  partitioning.     

\begin{figure}
\centering
\includegraphics[width=0.45\textwidth]{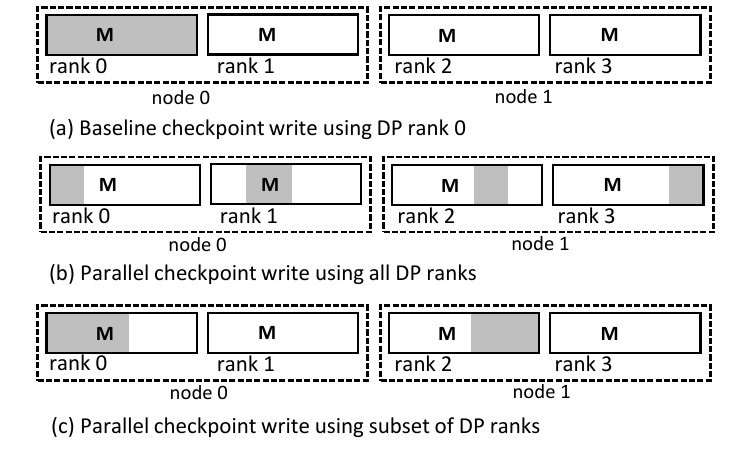}
\caption{\label{fig:dp_write}\emph Training model \textbf{M} on 2 nodes with DP=4. While (a) baseline uses only rank 0 to write checkpoints (shaded box), \name{} parallelizes checkpoint writes (shaded slices) across either (b) four DP ranks or (c) two DP ranks.}
\end{figure}

\textbf{Hardware efficiency:} Since DP training does not change checkpoint size, we consider two situations where using all the DP ranks to parallelize checkpoint creation could prove sub-optimal: (i) the checkpoint partition assigned to each rank is reduced to a size that prevents efficient disk write,  and (ii) contention for shared I/O hardware (e.g., PCIe or SSD) by the ranks is increased to a magnitude that hurts hardware efficiency. In these cases, checkpoint write bandwidth improves sub-linearly with respect to DP because of reduction in average bandwidth of ranks (or devices). In contrast, using a subset of the DP ranks, to increase write size per rank and reduce hardware contention, could yield better performance. Thus, in order to ensure good performance across many usage scenarios, \name{} provides the flexibility of using a subset of DP ranks to parallelize checkpoint creation. 

We illustrate the flexibility of \name{} checkpointing parallelism in Figure~\ref{fig:dp_write} for model {\it M} trained on two nodes with DP=4. In the baseline approach, Figure~\ref{fig:dp_write}(a), the model checkpoint is created by {\it rank 0} using the I/O hardware of {\it node 0}. In contrast, \name{} parallelizes checkpoint creation using I/O hardware of both nodes, and with the flexibility of using four ranks (Figure~\ref{fig:dp_write}(b)) or two ranks (Figure~\ref{fig:dp_write}(c)). In this example, parallelizing with two ranks instead of four increases write size per rank, and avoids I/O hardware contention. 

\name{} avoids selecting an arbitrary subset of DP ranks for checkpoint parallelism because some subsets hurt performance compared to using all DP ranks. Consider Figure~\ref{fig:dp_write}, parallel checkpointing using two ranks from the same node (i.e., {\it rank 0} \& {\it rank 1}, or {\it rank 2} \& {\it rank 3}) is worse than using all four ranks because of under-utilization of I/O hardware. Thus, \name{} chooses a subset of DP ranks that maximizes the utilization of, but minimizes contention for I/O hardware.

\textbf{Load balancing:} The partitioning algorithm must ensure that the checkpointing load is evenly balanced among the participating DP ranks to avoid {\it straggler effects} and resulting performance problems. Since our target are homogeneous hardware clusters where ranks have identical write bandwidth (GB/sec), the load of each rank is the size in bytes of the partition assigned to it. This requirement makes the simple approach of dividing the model layers among the DP ranks an unsuitable solution, because it would result in load imbalance for models with different types of layers. For example, large language models consist of different layer types (e.g., embedding, transformer, normalization, etc.), and the count and size (bytes) of these types are significantly different. Dividing the tensors among the ranks is similarly unsuitable.

\name{} partitions data on byte granularity to achieve load balancing among the parallel checkpoint writers. This approach depends on the actual number of bytes written out to disk, and so partitioning is done after tensor serialization and other in-memory processing that modify checkpoint size. Delaying the partitioning step like this helps to tightly bound imbalance to at most one byte. Loading parallel checkpoints is a two step process where each DP rank: (i) loads its checkpoint partition, if any, into GPU memory, and (ii) performs an {\it allgather} collective communication with other DP ranks of the model slice to assemble the full checkpoint state.

\subsection{Pipelining Checkpoint Writes}
\label{sec:pipeline-write}
The third optimization overlaps checkpoint creation with independent operations of the next iteration in order to reduce or avoid training stalls. As shown in Figure~\ref{fig:system_overview}(d), this optimizations improves performance by allowing the {\it forward} and {\it backward} passes of the next iteration to run immediately after the {\it optimizer} pass, and concurrently with the checkpoint creation. The {\it optimizer} pass of the next iteration is stalled only if checkpoint creation is not completed at the time, otherwise it proceeds with updating the model and training runs entirely without stalling. Synchronizing checkpointing and {\it optimizer} in this fashion is sufficient to satisfy the correctness requirement of the data dependency. 

For each training rank, this optimization introduces sharing of GPU compute and memory between a main thread and a helper thread. The main thread handles the compute (i.e., {\it forward}, {\it backward}, and {\it optimizer}) and communication (e.g., gradient reduction) aspects of distributed training, while the helper thread handles checkpoint creation. 

The two threads cooperate in the following fashion. The helper thread executes an infinite loop where it blocks until woken by the main thread to create checkpoints. The helper thread then proceeds to write the relevant tensors to persistent storage, signals completion to the main thread, and then blocks until the next request. The main thread blocks before {\it optimizer} to receive confirmation of the completion of the previous checkpoint creation request, and sends a new checkpoint creation request to the helper thread after {\it optimizer}. 

Since the main thread handles the critical path of training, we use the following design choices for the helper thread to reduce contention for GPU compute and memory. First, the helper thread's use of GPU memory is restricted to reading existing tensors, and so it does not allocate GPU memory. Second, the helper thread reads GPU tensors into page-locked CPU memory which uses DMA transfers and minimizes GPU cycles. Third, our write parallelization scheme is communication-free (\S~\ref{subsec:parallel_write}), so the helper thread does not need to execute GPU communication kernels. In summary, by ensuring that the helper thread mainly consumes CPU and I/O resources, we reduce contention with the GPU-bound main thread, and thus improve overall efficiency. 

    % \item Decouple checkpoint process from compute process
    % \begin{itemize}
    %     \item figure draw F,B,O on compute process, signalling to checkpoint process after O, etc
    % \end{itemize}

% \begin{itemize}
%      \item NVMe acceleration of single rank writes
%     \begin{itemize}
%         \item Checkpoint size may violate SSD block alignment (512B) = Use two file descriptors
%         \item Tensor-by-tensor writes = Write streaming to extend checkpoint file
%         \item Multi-hop write path from GPU/CPU memory to storage = Double buffering to overlap
%     \end{itemize}
%     %\item Efficient scaling through parallel writes
%     %\begin{itemize}
%         %\item Policy to split model into even segments = byte granularity splitting
%         %\item Scale with DP - split among model replicas
%         %\item Scale with hardware resouce (CPU socket, machine, SSD)
%     %\end{itemize}
%     \item Specialized optimizations
%     \begin{itemize}
%         \item Multi-socket machines 
%     \end{itemize}
% \end{itemize}
\section{Evaluation}
%We now present our evaluation of \name{} using real world dense and sparse DL models. 
We now evaluate the performance of \name{} for DL model training in the challenging scenario of checkpointing in every iteration. Our results are organized in the following way. First, we study the effectiveness our NVMe and parallelism optimizations for reducing checkpointing latency. For this, we use micro-benchmarks on single GPU and multi-node environments to measure checkpointing throughput (\S~\ref{subsec:micro_benchmark}), and real world dense and sparse DL models to additionally measure training speedups relative to baseline (\S~\ref{subsec:dense_models}, \S~\ref{subsec:sparse_models}). Second, we study the effectiveness of our decoupling strategy to hide checkpointing latency to achieve negligible training slowdowns (\S~\ref{subsec:eval_pipeline}). Finally, we estimate \name{} performance for DP larger than our GPU availability (\S~\ref{subsec:dp_projection}).  

\subsection{Implementation.}
%To simplify the integration of \name{} into the DL ecosystem we use two popular frameworks, PyTorch~\cite{pytorch_osdi} and DeepSpeed~\cite{deepspeed_url} in our implementation.

We build \name{} on PyTorch~\cite{pytorch_osdi} and DeepSpeed~\cite{deepspeed_url}.

\textbf{Checkpoint creation:} \texttt{torch.save()} in PyTorch provides flexible interface that allows a destination file to be optionally specified as an object which implements I/O routines (e.g., \texttt{write()}). We exploit this flexibility  by implementing \name{} in a compatible object that we pass to \texttt{torch.save()}. % The \texttt{write()} method of object allows us to intercept and accelerate writes with \name{} optimizations. 
Our integration enables \texttt{torch.save()} to use \name{} for disk writes instead of standard I/O routines, with no change to other operations (e.g., tensor serialization).   

\textbf{NVMe optimizations:} DeepSpeed provides an NVMe-optimized module for fast tensor offloading~\cite{zero-infinity, deepspeed-inference}. We extend this module to support a file creation pattern involving multiple segment writes to increasing offset positions. This matches the standard way of creating checkpoints through multiple writes of serialized tensors (\S~\ref{subsubsec:background_state}).  

\textbf{Checkpoint pipelining:} The standard behavior of DeepSpeed is to create a single Python process for each training rank to perform training and checkpointing operations on the GPU. We extend DeepSpeed for checkpoint pipelining by creating a second and dedicated Python process for each rank for checkpointing, since multi-threading is inefficient due to python GIL.%~\footnote{A multi-threading approach would be inefficient because of Python GIL}. 
We leverage CUDA multiprocessing support to coordinate the training and checkpointing processes. 

% We implemented \name using checkpoint creation utility (i.e., \emph{torch.save}) of PyTorch~\cite{pytorch-osdi} v1.12 and the tensor offloading technology of DeepSpeed~\cite{zero-infinity}. 

% We inject XYZ into pytorch model checkpointing by replacing file object write procedure. More specifically, when we call \emph{torch.save(file\_objs, ...)}, we will save model layers as serialized file objects and write the file objects into disk. Therefore, instead of using PyTorch or Python's default file writer, we replace it with our own file writer to achieve faster object writes into local NVMe devices. In our implementation, we disable sort by key procedure in file objects serialization to get a consist view of tensor writing list during every training iteration. 

% To achieve faster NVMe write performance, when opening the file, we pass in \emph{O\_DIRECT} flag to reduce cache effects. We first hold the incoming file objects inside CPU memory to hide end-to-end model checkpointing latency. Once our CPU memory buffer is full, we flush the data from CPU memory buffer to local NVMe SSDs. We further reduce end-to-end model saving latency by adopting our double buffer design (described in Section 4.2) which achieves higher frequency of data flushing from CPU memory to NVMe devices.

\subsection{Methodology}

\subsubsection{Hardware}
\label{subsubsec:hardware}
We evaluated \name{} on a cluster consisting of $8$ DGX-2 machines connected via Infiniband. Each machine contains $16$ Nvidia V100 GPUs with 32GB on-device memory, for a total of $128$ GPUs. There are $8$ locally attached NVMe SSDs on each machine, configured into a single RAID-0 volume, a combined peak write bandwidth of $24.8$ GB/sec. 

\subsubsection{Models}
\label{subsubsec:models}

\begin{table}
\small
 \caption{Experiment details of GPT-3 models.}
\begin{center}
\label{tb:gpt-setup}
\begin{tabular}{c|c|c|c|c} 
 \hline
  Model & Dense & Model/Expert & Global  & Checkpoint\\ 
  Size  &       & Parallelism & Batch Size & Size (GB) \\
\hline
0.7B & Yes & 1 & 256 & 10 \\
\hline
1.3B & Yes & 2 & 512 & 17 \\ 
%\cline{2-3}
\hline
2.7B & Yes & 4 & 512 & 35 \\ 
\hline
6.7B & Yes & 8 & 1024 & 88 \\ 
\hline
13B & Yes & 16 & 1024 & 173 \\ 
\hline
1.8B-MoE & No & 16 & 256 & 67 \\ 
%\cline{2-6}
 \hline
\end{tabular}
\end{center}
\end{table}

For our experiments, we use the dense and sparse DL models listed in Table~\ref{tb:gpt-setup}, all of which are based on the GPT-3~\cite{gpt3} model architecture. The five dense models range in size from $0.7$ to $13$ billion parameters, and are configured, in terms of model parallelism (MP), and global batch size (GBS), based on details in~\cite{gpt3}. Dense models with $MP>1$ use only tensor parallelism (TP), except {\it gpt3-13B} which also uses pipeline parallelism (PP) (i.e., TP=$8$, PP=$2$). The sparse model is a $1.8$ billion parameter Mixture of Experts (MoE), with expert parallelism (EP) degree of $16$ and GBS of $256$, based on details in~\cite{deepspeed-moe}. Table~\ref{tb:gpt-setup} reports the checkpoint sizes. We use gradient accumulation (GA) to align our experiments with ~\cite{gpt3, deepspeed-moe} due to the data parallelism (DP) limits of our cluster.  

\iffalse
The model architecture we select in experiments is GPT-3 series in both dense and MoE veriations (see Table ~\ref{tb:gpt-setup}). As illustrated in Table~\ref{tb:gpt-setup}, we distribute and train these giant DL models via data parallelism, tensor parallelism and pipeline parallelism jointly. We form data parallel groups using different number of GPUs for different model sizes. For GPT 0.7B, we adopt pure data parallelism, where each GPU holds the full copy of model parameters. For GPT 1.3B, for each data parallel group, we split model over 2 GPUs with tensor parallelism. For GPT 2.7B/6.7B, we spread each model across 4/8 GPUs with tensor parallelism, respectively. For GPT 13B, we use 16 GPUs to hold one model parameters with 8 tensor parallelism and 2 pipeline parallelism. For 1.8B MoE model, we spread its 16 experts of 0.8B on to 16 GPUs directly.

Note that our checkpointing methodology can only be parallelized in data parallelism dimension, we report our scaling results in data parallelism dimension. 
\fi 

\begin{figure}[t]
    \centering
    \subfigure[Single GPU, \textit{16MB Checkpoint}]{\includegraphics[width=0.23\textwidth]{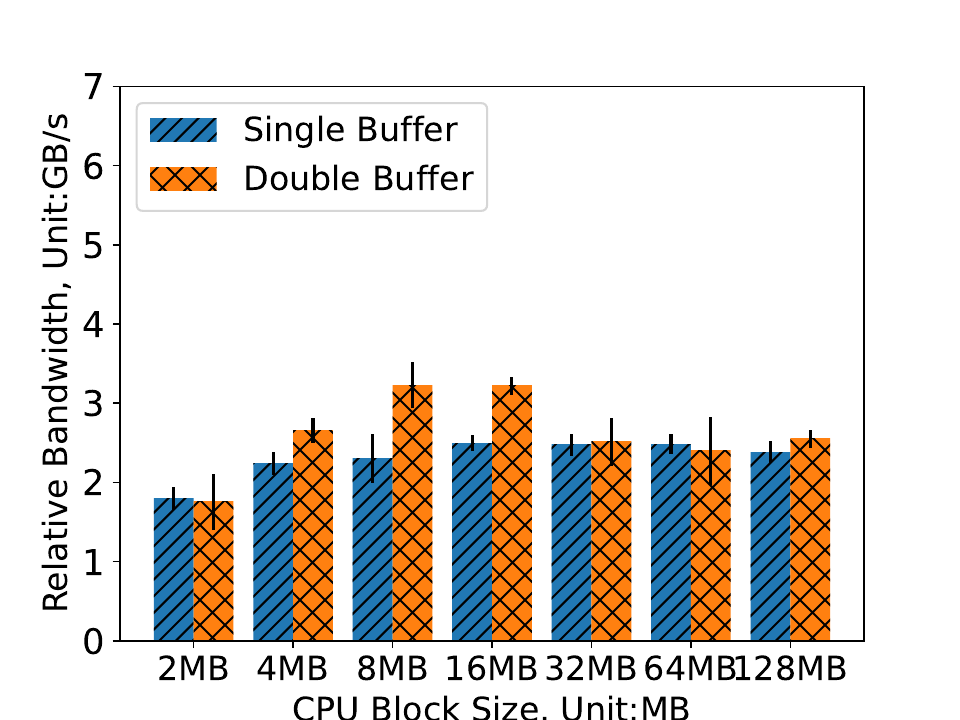}} 
    \subfigure[Single GPU, \textit{512MB Checkpoint}]{\includegraphics[width=0.23\textwidth]{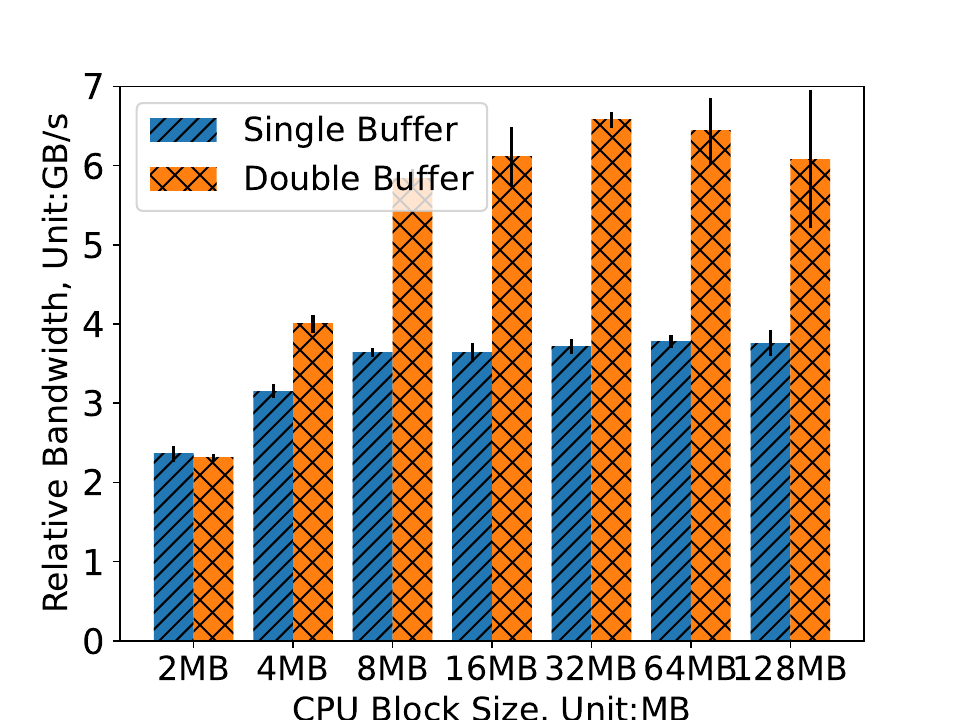}} 
    \caption{Varying {\it IO Buffer} size on single GPU}
    \label{fig:block_size}
%\vspace{-6mm}
\end{figure}

\begin{figure}[t]
    \centering
    \subfigure[\textit{2 Nodes (4 Sockets).}]{\includegraphics[width=0.23\textwidth]{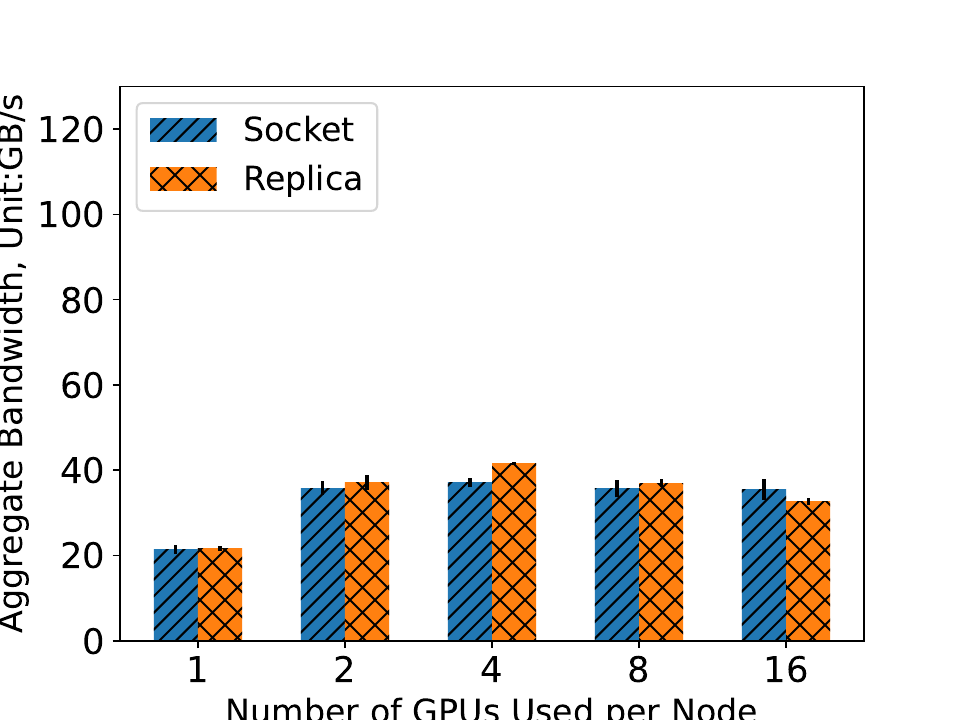}} 
    \subfigure[\textit{8 Nodes (16 Sockets).}]{\includegraphics[width=0.23\textwidth]{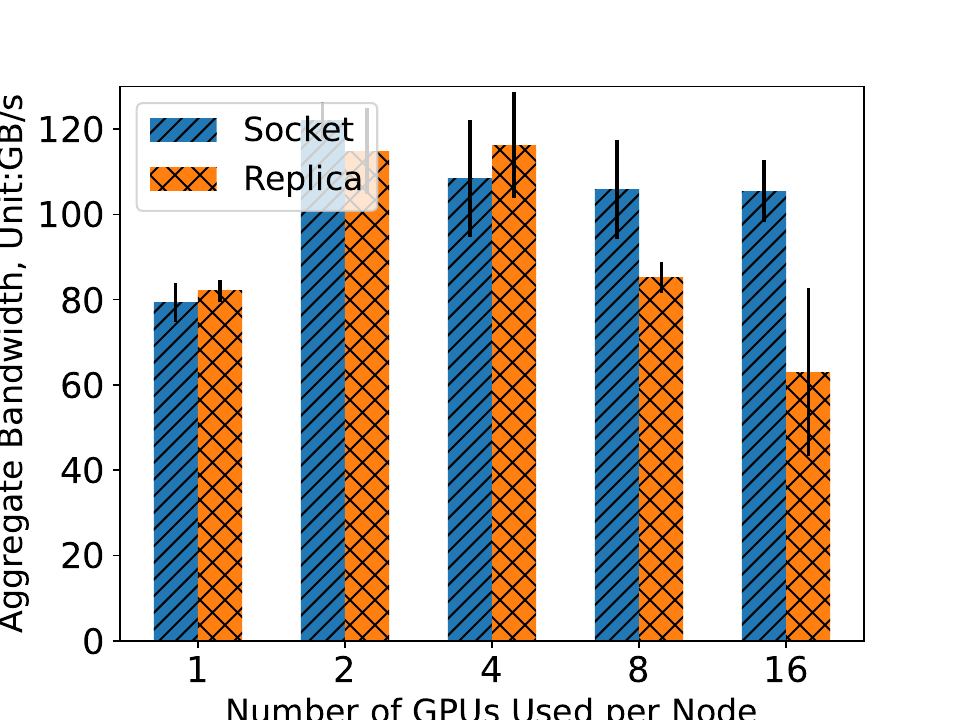}} 
    \caption{Parallel checkpointing of {\it gpt3-0.7b}.}
    \label{fig:multi_nodes}
\end{figure}

\subsection{Micro Benchmark}
\label{subsec:micro_benchmark}
Here we evaluate the proposed schemes of overlapped checkpoint writes  (\S\ref{subsec:accel_write}) and write parallelism (\S\ref{subsec:parallel_write}). %In particular, to evaluate the proposals under various checkpointing scenarios and system conditions, we design two sets of experiments to obtain a good coverage of checkpoint sizes, system configurations, and training scales and environments. 

\subsubsection{Single GPU Experiments}
A core aspect of our NVMe optimizations is using page-locked CPU memory (a.k.a., {\it IO Buffer}) to write checkpoint data from accelerator memory to SSD. In this set of single GPU experiments we study four aspects of these optimizations: (i) speedup over baseline \texttt{torch.save()}, (ii) checkpoint data size, (iii) {\it IO Buffer} size, and (iv) single buffer and double buffer write modes. We measure the bandwidth of writing GPU tensors of sizes ranging from $16$MB to $512$MB to SSD. We chose this size range to represent the per rank checkpoint load in training scenarios where \name{} parallelizes GBs/TBs of checkpoint data over dozens/hundreds of DP ranks. In Figure~\ref{fig:block_size}, we report the relative bandwidth of \name{} over baseline for $16$MB and $512$MB checkpoints and {\it IO Buffer} sizes of $2$MB $\sim$ $128$MB. We see similar results for $32$MB $\sim$ $256$MB checkpoints but leave those to appendix.% to save space.  

% In the first set of experiments, we focus on the performance of the page locked CPU memory 
% In the first set of experiments, we focus on the performance of the default Pytorch checkpointing scheme (\texttt{torch.save()}) and the single and double buffer schemes on various checkpoint sizes and various CPU block sizes on a single CPU and GPU from a single node. 
% We focus on single GPU execution while varying checkpoint size from 16MB to 512MB, and {\it IO Buffer} size from 2MB to 128MB. We choose the checkpoint sizes to reflect the wide ranges of training states from GBs to TBs and training scales from several CPUs/GPUs to hundreds of millions of CPUs/GPUs.  We use the performance of \texttt{torch.save()} on per checkpoint size as the baseline, and report the relative bandwidths of our single and buffer schemes accordingly.  Figure~\ref{fig:block_size} presents the results of 16MB and 512 MB checkpoints on GPU. We see the similar trends and patterns from the results of 32MB --- 256MB on GPU and all of the results on CPU and leave all of these results to the appendix due to the similarity and space issue.  

First, we see that \name{}, in either single or double buffer mode, consistently outperforms \texttt{torch.save()} . For all checkpoint sizes, we observe performance improvements of 1.8X $\sim$ 3.6X, and 1.8X $\sim$ 6.6X for single and double buffer modes respectively. These results show significant efficiency benefits of \name{}, even on a single GPU.

% First, Figure~\ref{fig:block_size} shows clearly that, the single and double buffer schemes consistently deliver better performance than \texttt{torch.save()} does. In particular, across checkpoint sizes and CPU block sizes, the two schemes obtain 1.8X --- 3.64X (single buffer scheme) and 1.75X --- 6.58X (double buffer scheme) performance improvement respectively. These observations suggest that our proposed schemes work efficiently and outperform the Pytorch default proposal significantly.   

Second, we see that the benefits of \name{} increases with checkpoint data size. For single buffer mode, the maximum improvement increases from $2.5$X for $16$MB checkpoints to $3.6$X for $512$MB. Similarly, for double buffer mode, the maximum improvement increases from $3.6$X for $16$MB to $6.6$X for $512$MB. While it is known that disk write efficiency improves with size, these results show that \name{} achieves better efficiency improvement compared to baseline.  

Third, we see that {\it IO Buffer} size affects both single and double buffer modes, and in different ways for different checkpoint sizes. For 16MB checkpoints, the lowest performance of both modes is with 2MB {\it IO Buffer}, while the best performance is with 16MB (single buffer) and 8MB (double buffer). Also, the performance ratio of the best over the worst is 1.38X (single buffer) and 1.43X (double buffer). Similarly, for 512MB checkpoints, the lowest performance of both modes with 2MB {\it IO Buffer}, and the best performance is with 64MB (single buffer) and 32MB (double buffer). In this case, the performance ratio of best over the worst is 1.6X (single buffer) and 2.87X (double buffer). Thus, careful {\it IO Buffer} size configuration is rewarding, especially for large checkpoints.   

% Second, it is clear that, both of the single and double buffer schemes are sensitive to CPU block size, and different checkpoint sizes benefit from different block sizes. In particular, for 16MB checkpoints, the two schemes deliver the worst relative mean performance both with 2MB block size, and deliver the best relative mean performance when with 16MB and 8MB block sizes respectively. Moreover, for the two schemes, the ratio of the best over the worst is 1.38X (single buffer scheme) and 1.43X (double buffer scheme) respectively. 
% Similarly, for 512MB checkpoints, the two schemes deliver the worst performance when both with 2MB block size, and deliver the best performance when with 64MB and 32MB block sizes respectively. In this scenario, the ratio of the best over the worst is 1.6X (single buffer scheme) and 2.87X (double buffer scheme) respectively. Clearly, the feature of configurable CPU block size is useful, especially for large checkpoints for whom appropriate configuration can improve performance efficiently.   

Fourth, double buffer mode is faster than single buffer mode for most {\it IO Buffer} and checkpoint data sizes. For the studied seven {\it IO Buffer} sizes, double buffer mode is better in five cases for 16MB checkpoints (1.01X $\sim$ 1.4X faster), and in six cases for 512MB checkpoints (1.2X $\sim$ 1.77X faster). In the few cases where single buffer mode is better, the performance gap is $\le$3\%. Moreover, double buffer mode obtains the best performance for both 16MB (5.18GB/s) and 512MB (10.9GB/s) checkpoints. Thus, double buffer mode is generally preferable, especially for well configured {\it IO Buffer}.

% Third, across checkpoint sizes and CPU block sizes, double buffer scheme delivers better performance in most cases, and consistently deliver the best performance for various checkpoint sizes. In particular, for 16MB and 512MB checkpoints, double buffer scheme outperforms single 
% buffer scheme in 71.4\% (5 out of 7) of the block-size configurations for 16MB checkpoints and in 85.7\% (6 out of 7) of the configurations for 512MB checkpoints, with the outperformance ranging in 1.01X --- 1.4X (16MB checkpoints) and in 1.2X --- 1.77X (512MB checkpints) respectively. In the cases where single buffer scheme performs better, the outperformance is $\le$3\%. Moreover, double buffer scheme obtains the best performance for both 16MB (5.18GB/s) and 512MB (10.9GB/s) checkpoints. This suggests that double buffer scheme is generally a better choice and can consistently deliver the best performance when CPU block sizes are appropriately configured.    

\subsubsection{Multi-Node Experiments}

For DP training, \name{} can parallelize checkpoint writes using all or some DP ranks. In this set of experiments, we study the performance of these two options by measuring the bandwidth of writing a {\it gpt3-0.7b} checkpoint ($\sim 10$GB) from GPU memory to SSD. Each DGX-2 node has two CPU sockets, and so in our experiments, we configure write parallelism with some DP ranks as one writer per CPU socket for higher CPU and PCIe utilization. In our results, we label using all DP ranks as {\it Replica}, and using some DP ranks as {\it Socket}. We collect results on up to $8$ nodes, using up to $16$ GPUs per node. Figure~\ref{fig:multi_nodes} presents the results of $2$ and $8$ nodes. We see the similar patterns in the results of 1 and 4 nodes, and leave their results to the appendix to save space.

% In the second set of experiments, we evaluate the socket and replica parallelisms across training scales. We train and checkpoint GPT3-large with the proposed two parallelisms on various number of nodes and various number of GPUs per node. In particular, we vary the number of nodes in 1 --- 8 nodes and vary the number of GPUs per node in 1--- 16 GPUs. Figure~\ref{fig:multi_nodes} presents the results of 2 and 8 nodes. We see the similar patterns from the results of 1 and 4 nodes, and leave their results to the appendix due to the similarity and space issue.

We observe that the relative performance of {\it Replica} and {\it Socket} depends on interaction of the benefits of increasing write parallelism degree, increased available bandwidth, and the downsides, increased I/O hardware contention and reduced per rank write size. From Figure~\ref{fig:multi_nodes}, we see that the best write parallelism degree is $8$ and $16$ for $2$ and $8$ nodes respectively. On $2$ nodes, this corresponds to $4$ GPUs per node, is possible only with {\it Replica} ({\it Socket} is restricted to $4$), and achieves $41.8$GB/sec write bandwidth ($91\%$ of SSD peak). On $8$ nodes, this corresponds to $2$ GPUs per node, is possible with both {\it Replica} and {\it Socket}, but {\it Socket} achieves the best aggregate bandwidth of $129.8$GB/sec ($87\%$ of SSD peak). We observe that scaling beyond these parallelism degrees degrades {\it Replica} performance (especially on $8$ nodes) due to the aforementioned downsides. These results show the benefit of {\it Socket} write parallelism for large-scale DP training.

\subsection{Real World Dense Models}
\label{subsec:dense_models}
We study the performance benefits of \name{} for dense models in terms of checkpointing and end-to-end training. 

%We study how \name{} performs on dense model training. We report write speedup and end-to-end training speedup we achieves by adopting \name{} over the baseline. %We also disucss end-to-end model training speedups for model checkpointing at each training iteration.

\begin{figure*}[t]
    \centering
    \subfigure[Checkpointing speedup]{\includegraphics[width=0.24\textwidth]{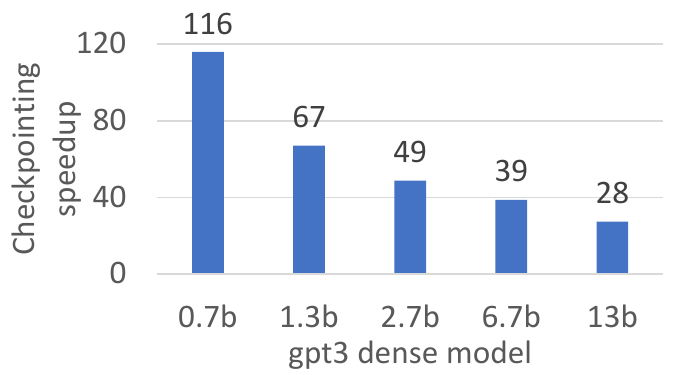}\label{fig:dense_model_ckpt_speedup}} 
    \subfigure[Checkpoint throughput scaling]{\includegraphics[width=0.24\textwidth]{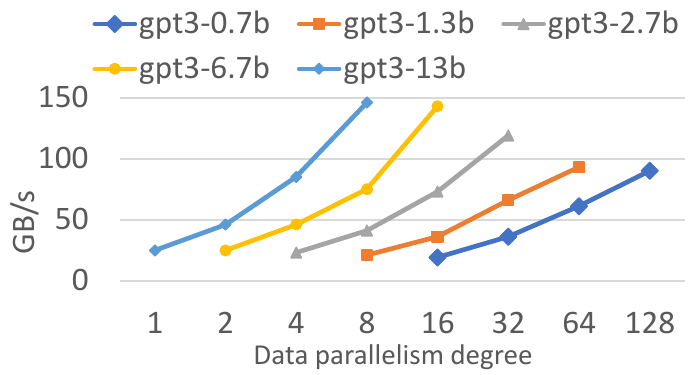}\label{fig:dp_scaling_dense_ckpt}} 
    \subfigure[Training speedup on 8 nodes]{\includegraphics[width=0.24\textwidth]{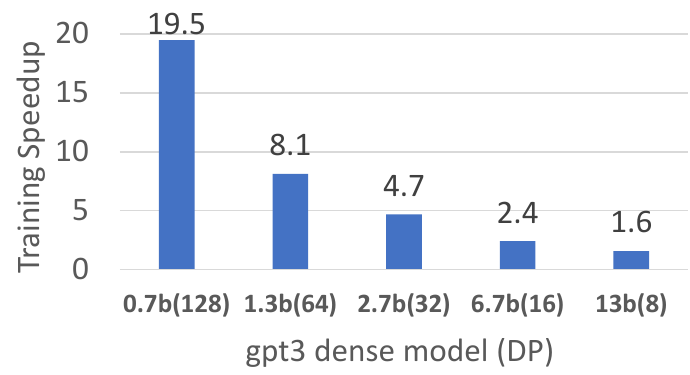}\label{fig:dense_model_training_speedup}} 
    \subfigure[Training speedup relative to DP]{\includegraphics[width=0.24\textwidth]{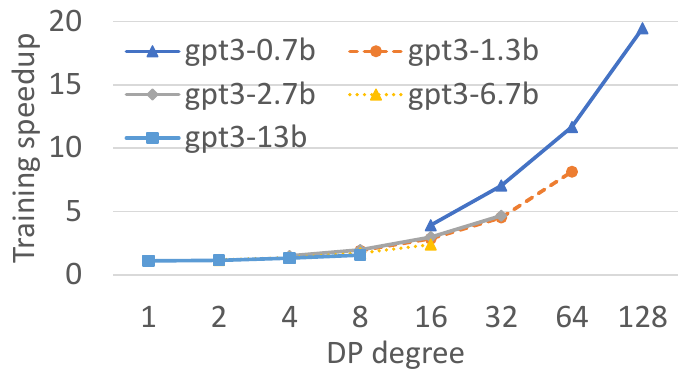}\label{fig:dp_scaling_dense_training}} 
    \caption{\name{} on GPT3 dense model training.}
    \label{fig:dense_model_perf}
\end{figure*}

\subsubsection{Checkpoint Speedup}
We observe that \name{} creates checkpoints significantly faster than baseline. Figure~\ref{fig:dense_model_ckpt_speedup} shows that on $128$ V100 GPUs, \name{} achieves speedups ranging from $28$x ({\it gpt3-13B}) to $116$x ({\it gpt3-0.7B}). These improvements demonstrate the effectiveness of our NVMe and parallelism optimizations. Speedup decreases as model size increases because larger models have smaller DP degree, for a fixed GPU count, and thus lower checkpointing parallelism. For example, DP degrees of {\it gpt3-0.7B} and {\it gpt3-13B} are $128$ and $8$ respectively.  

%Figure~\ref{fig:dense_model_ckpt_speedup} depicts the speed up numbers of \name{} model checkpointing over the baseline using 128 V100 GPUs. With pure data parallel training like GPT-3 0.7B, we achieve 116x speedup over baseline, which is near linear scalability. The main reason of not achieving $N$x speedup over $N$ DP ranks is due to contention on shared PCIe and NVMe devices. For tensor parallel training on dense model like GPT-3 1.3B, we achieves 67x data persistence speed up over baseline using 64 DP groups, which is over linear scalability. The main reason here is \texttt{torch.save()} use traditional file I/O library as the file write backend, which cannot fully utilize the write bandwidth on NVMe devices. Similar results can be found in for training larger models using both pipeline parallelism and tensor parallelism. For GPT-3 2.7b, 6.7b, we achieve 49x/39x checkpoint write speedup over baseline with 32/16 DP groups. The biggest gap we achieve over linear scalability is on GPT-3 13B model. We reach 28x speedup over baseline with 8 DP groups, which is 3.5x (28/8) over linear scalability. 

Figure~\ref{fig:dp_scaling_dense_ckpt} reports the checkpointing throughput of \name{} for the configurable DP degrees of each model on up to $128$ GPUs. We make three observations: (i) \name{} throughput scales with DP degree for all models, (ii) \name{} writes checkpoints at up to $146$GB/sec ({\it gpt3-13b}), which is $80\%$ of theoretical peak on $8$ nodes, and (iii) for a DP degree, \name{} is more efficient on larger models due to larger writes per parallel writer.    

%Figure~\ref{fig:dp_scaling_dense_ckpt} shows throughput we achieve when scaling out to different degree of DP using up to 128 GPUs. In Figure~\ref{fig:dp_scaling_dense_ckpt}, for larger models like GPT3-13b and GPT3-6.7b, we can reach around 150 GB/s of aggregated checkpoint write throughput. For medium size model like GPT3-2.7b, we achieve up to around 120 GB/s file write speed. Even for small model like GPT3-0.7b and 1.3b, we can reach around 100 GB/s using parallel writers due to our decent load balancing.  

%Figure~\ref{fig:dense_model_perf}(a) and ~\ref{fig:dense_model_perf}(b). 

\subsubsection{E2E Training Speedup}
We study the speedup provided \name{} over baseline for end-to-end training with checkpointing on every iteration. Figure~\ref{fig:dense_model_training_speedup} reports that for training on $128$ GPUs, \name{} speedups are in the range of $1.6$x ({\it gpt3-13B}) to $21.8$x ({\it gpt3-0.7B)}. As explained earlier, we observe higher speedups for smaller models (e.g., {\it gpt3-0.7B} because their higher DP degree enables higher write parallelism. This speedup trend is better illustrated by Figure~\ref{fig:dp_scaling_dense_training}, which reports training speedups as a function of DP degree. We observe that for a given DP degree, \name{} achieves similar speedups for all the models. However, since \name{} speedup increases with DP degree, smaller models (e.g., {\it gpt3-0.7B)} enjoy higher speedups as they can be trained with higher DP degrees in our cluster. We expect that the speedups of larger models (e.g., {\it gpt3-13B}) will increase with DP, and we test this conjecture in \S~\ref{subsec:dp_projection} by projecting DP degree to $128$.

\iffalse 
We compare the performance of using \name{} for checkpointing 
Figure~\ref{fig:dense_model_training_speedup} reports end-to-end model training speedup over baseline. Here we measure the total iteration time including both model training time and checkpointing time using 8 nodes (128 GPUs in total). Our biggest win here is in high DP degree scenarios. For model training of GPT-3 0.7b with DP degree of 128, we achieve 19.5x speedup of training iteration time over baseline. The less DP degree we have, the less speedup we may achieve. For model size ranging from 1.3b to 13b, our iteration time speedup decrease from 8.1x to 1.6x. The main reason for less speedup in larger model is due to limited DP degree. We project our training speedup on large models like GPT-3 13B using 1024 GPUs in Section 5.8.

Figure~\ref{fig:dp_scaling_dense_training} shows the impact of varied DP degree on our end-to-end training speedup results. In a nutshell, the higher DP degree leads to high end-to-end training speedup numbers. The key insight is computation time decreases proportionally to DP degree while model checkpointing time remains the same for each training iteration. 
\fi 

\subsection{Real World Sparse Models}
\label{subsec:sparse_models}
Sparse models have recently gained attention as way to scale model size without increasing the computation requirements~\cite{gshard, switch, deepspeed-moe}. We study the performance benefits of \name{} for sparse models in terms of checkpointing and end-to-end training. We use {\it gpt3-1.8B-MoE} with EP=$16$, which means a model replica occurs a node and restricts DP to $\le 8$. 

%We further extend evaluations to Mix of Experts (MoE) sparse models. Here we mainly target on the MoE model of gpt3-0.3b with 16 experts. Since 16 GPUs form a DP group, we evaluate our performance with DP degree varied from 1 to 8. 

\begin{figure}[t]
    \centering
    \subfigure[Speedups.]{\includegraphics[width=0.23\textwidth]{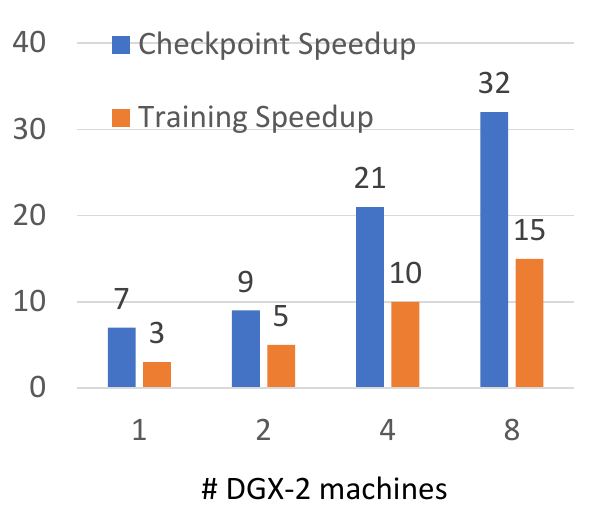}\label{fig:moe_speedup}} 
    \subfigure[Throughput scaling with DP.]{\includegraphics[width=0.23\textwidth]{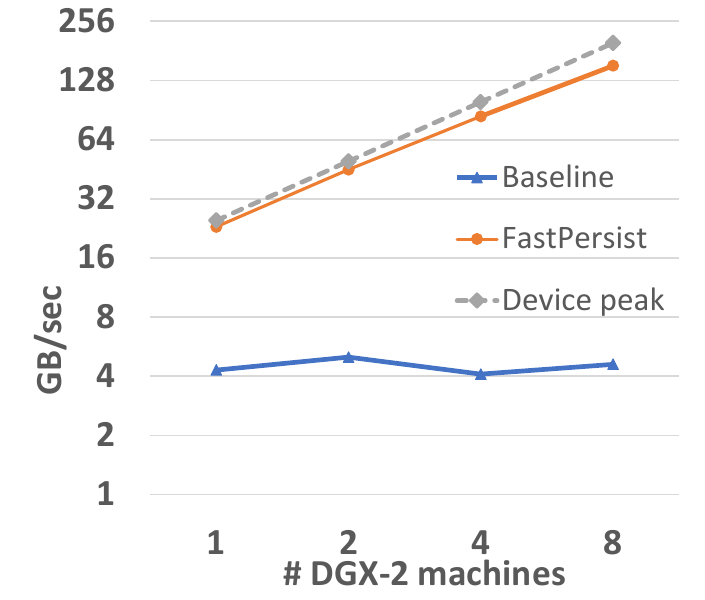}\label{fig:dp_scaling_moe_ckpt}} 
    %\subfigure[Training speedup.]{\includegraphics[width=0.3\textwidth]{figures/moe_training_speedup.png}\label{fig:moe_training_speedup}} 
    \caption{\name{} on {\it gpt3-1.8B-MoE}.}
    \label{fig:moe_perf}
\end{figure}

\subsubsection{Checkpointing Speedup}

Similar as Section 5.4.1, blue bars in Figures~\ref{fig:moe_speedup} shows our speedup over baseline on MoE model. Compared with our results on dense models, we achieve higher speedup on sparse models with the same DP degree. For example, compared with the dense counterpart GPT-3 13b which also use 16 GPUs per DP group, we can achieve 32x speedup in MoE models with 8 DP degree whereas GPT-3 13b achieve 28x speedup in Figure~\ref{fig:dense_model_ckpt_speedup}. Furthermore, we can even achieve 7x speedup with just DP=1. This is mainly because sparse models need to checkpoint more data compared with dense models. 

In Figure~\ref{fig:dp_scaling_moe_ckpt}, we show our MoE checkpoint performance over DP scaling. Baseline performs pooly in this case, which only achieve around 4GB/s writing throughput. As shown in Figure~\ref{fig:dp_scaling_moe_ckpt}, we achieve near linear scalability from 1 node to 8 nodes and very close to hardware upperbound, which also verifies the effectiveness of our implementation. 

\subsubsection{E2E Training Speedup}
For orange bars in Figure~\ref{fig:moe_speedup}, we report training time speedup for MoE model. Compared with results  in Figure~\ref{fig:dp_scaling_dense_training}, here we achieve higher speedup with the same DP degree. For example, the iteration time is less than 2x given DP degree of 8 in dense models in Figure ~\ref{fig:dp_scaling_dense_training}. In contrast, in MoE case as orange bars in Figure~\ref{fig:moe_speedup}, we can achieve 15x end-to-end iteration time speedup with same DP degree of 8. It shows the trend that \name{}'s performance can be more pronounced or amplified in sparse model scenarios. 

\begin{figure}[t]
    \centering
    \subfigure[Impact of DP and GAS]{\includegraphics[width=0.23\textwidth]{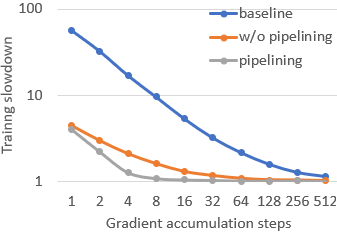}\label{fig:dense_1_3b_pipeline_gas_dp}} 
    \subfigure[Training slowdown on 8 nodes]{\includegraphics[width=0.24\textwidth]{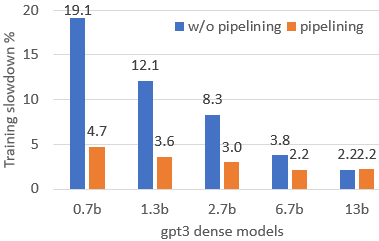}\label{fig:dense_model_pipeline_slowdown}} 
    \caption{\name{} pipelining performance.}
    \label{fig:pipeline_perf}
\end{figure}

\subsection{Pipelined Checkpointing}
\label{subsec:eval_pipeline}
%As describted in \S\ref{sec:pipeline-write}, we further reduce the checkpoint overhead by decoupling the checkpoint write process from the  training process. 
We now evaluate the additional benefits of using pipelining to reduce checkpointing stalls (\S\ref{sec:pipeline-write}). 

\subsubsection{Sensitivity Analysis}
We run {\it gpt3-1.3B} with DP=$1$ and compare the training slowdown of checkpointing on each iteration for baseline and \name{} with and without pipelining. We use GPU $0$ and $8$ to minimize DRAM and PCIe bandwidth interference. As discussed in \S\ref{subsubsec:background_batch}, higher GAS values (lower DP values) increases compute time and lowers checkpointing overhead.  So, we sweep over gradient accumulation step (GAS) $1$ to $512$ to observe the interaction of GAS and DP on checkpointing performance. The results summarized in Figure~\ref{fig:dense_1_3b_pipeline_gas_dp}, show that pipelining is better for GAS $<64$, and achieves negligible slowdown earlier ($8\%$ at GAS=$8$).

\begin{figure}
\centering
\includegraphics[width=0.35\textwidth]{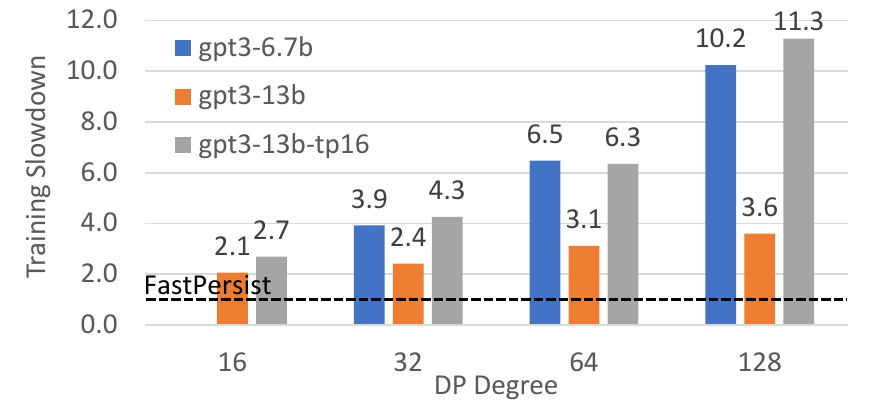}
%\vspace{-3mm}
\caption{\label{fig:128-sim} Projecting to DP $\le 128$.}
%\vspace{-3mm}
\end{figure}

\subsubsection{Training Overheads of Frequent Checkpointing}
We use the dense GPT3 models to evaluate the effectiveness of \name{} in reducing the training overheads of checkpointing on every iteration. We fix the number of nodes to $8$ and measure \name{} slowdown with and without pipelined checkpointing. In Figure~\ref{fig:dense_model_pipeline_slowdown}, the results show that checkpoint pipelining consistently provides additional benefits to checkpoint acceleration. Moreover, we observe that for model size between 1.3b to 13b, we can achieve $<5\%$ overhead when checkpointing at every training iteration, which can be negligible. It indicates that we can checkpoint large NLP models at every iteration for almost free, which is quite impressive.

\subsection{Projection to Larger Scale}
\label{subsec:dp_projection}
Given GPU hardware constrain, we simulate performance results for big dense models like GPT-3 6.7B and 13B by projecting up to 128 DP degree (i.e., 1024 GPUs for 6.7B, 2048 GPUs for 13B). For 6.7B with DP degree of 16, we omit simulation since it is evaluated in Figure~\ref{fig:dense_model_training_speedup}.

Figure~\ref{fig:128-sim} shows the projected training speedup numbers that \name{} achieves over the baseline, where blue/orange bars are for 6.7B/13B model respectively. When scaling out to thousands of GPUs, \name{} maintains consistent checkpointing overhead ($<2\%$ of training computation time) whereas baseline checkpointing overheads grows proportionally to DP degree. For 6.7B and 13B models, we achieve up to 10.2x and 3.6x training speedup over baseline, separately.

To further reduce computation time for 13B model, we abandon PP and apply fully TP over 16 GPUs as a DP group. As grey bars in Figure~\ref{fig:128-sim}, \name{} achieves much higher speedup numbers over the baseline when comparing with standard TP and PP combined model split (i.e., orange bars in Figure~\ref{fig:128-sim}). We can achieve up to 11.3x training speedup over baseline in this full TP setting.

% \begin{itemize}
%     \item Implementation
%     \begin{itemize}
%        \item Torch checkpoint writer (torch.save())
%        \item DeepNVMe
%     \end{itemize}
%     \item Methodology
%     \begin{itemize}
%        \item Baseline: torch.save()  //Torch Snapshot, FSDP? 
%        \item Hardware: A100-80GB, V100-32GB
%        \item Model architectures: BERT, GPT, Turing, BLOOM
%        \item Training parallelism: DDP, TP, PP, 3D
%     \end{itemize}
%     \item Micro benchmark performance
%     \begin{itemize}
%         \item Metric: GB/sec and speedup
%         \item Workload:
%         \begin{itemize}
%             \item Write large tensors such as layers of large-scale models
%             \item Write model checkpoints
%         \end{itemize}
%         \item Single writer
%         \item Multiple writers (up to 16 nodes)
%     \end{itemize}
%     \item Real-world performance
%     \begin{itemize}
%         \item Fix global batch size to paper
%         \item Metric: E2E TFLOP/GPU
%         \item No. Nodes: 16
%         \item Workload: large model training
%     \end{itemize}
% \end{itemize}
\section{Related Work}
Checkpointing~\cite{ckpt-overview} has been studied in distributed systems with various goals~\cite{dist-snapshots, ckpt-recover, multi-level1, multi-level2, check-in, cog-ssd, HolisticGNN}. We summarize the related works into following three categories.

The first line of work is decoupling checkpointing from model training process~\cite{checkfreq, check-n-run}. Particularly, CheckFreq~\cite{checkfreq} conducts training workload profiling and automatically tunes checkpoint frequency given different software and hardware configurations. Moreover, it decouples checkpointing from the training pipeline. 
%Besides decouple checkpointing from model training process and data sharding and persistence in data parallel dimension,
Similarly, Check-N-Run~\cite{check-n-run} utilizes checkpointing decoupling and further reduces checkpoint size via quantization and only persists the changed values. These works complete model checkpointing in 2 steps: first persist data in volatile host memory and return, then flush data to non-volatile storage in the background. Different from these arts, \name{}'s single writer directly write data from GPU memory to non-volatile SSDs and achieve similar throughput as writing data to host memory. In addition, compared with quantization scheme in Check-n-run~\cite{check-n-run}, \name{} further reduces checkpoint overhead to be negligible by adopting parallel writers without losing data precision.

The second line for boosting-up checkpoint speed is model sharding and using parallel writers. The works like DeepFreeze~\cite{deepfreeze} achieve in-parallel checkpointing in data parallel dimension. However, it is only applicable for checkpointing of small models and focus on CPU cluster. In contrast, \name{} adopts similar approach but applies the parallelism on giant deep learning model checkpointing on GPUs. 

In addition, neither of the aforementioned works is specially optimized for NVMe storage. Besides deep learning workloads, \name{}'s optimization on NVMe device can be generally applicable to any data persistence process (e.g., data persistence in database, mapreduce systems).

%Third line: SSD for faster GPU workload (nvme)

%One line is to make checkpointing faster using alternative approaches~\cite{checkfreq, check-n-run, veloc}. To improve data persistence throughput in heterogeneous storage, Veloc~\cite{veloc} dynamically adjust asynchronous write concurrency to guarantee data always go to local fast devices first and never write to slow local disks directly. CheckFreq~\cite{checkfreq}, DeepFreeze~\cite{deepfreeze} and Check-n-run~\cite{check-n-run} are tailor made for deep learning workloads. 
%More specifically, CheckFreq~\cite{checkfreq} conducts training workload profiling and automatically tunes checkpoint frequency given different software and hardware configurations. To further reduce checkpoint latency overhead, it decouples model checkpointing from the training pipeline. DeepFreeze~\cite{deepfreeze} achieves in-parallel checkpointing in data parallel dimension. Besides decouple checkpointing from model training process and data sharding and persistence in data parallel dimension, Check-n-run~\cite{check-n-run} further reduces checkpoint size via quantization and only persists the changed values. However, none of the art above is specially optimized for model checkpointing on NVMe storage. \name{}'s optimization on NVMe device can be generally applicable to any data persistence process (e.g., data persistence in database, mapreduce systems). \name{} further reduces checkpoint overhead to be negligible by adopting parallel writers without losing data precision.  

The last line of related literature is to utilize SSD to accelerate GPU workloads~\cite{zero-infinity, spin2017}. Spin~\cite{spin2017} integrates peer-to-peer (p2p) access between disk and GPU into file I/O layer and activates p2p functionality when necessary, which achieves decent throughput for data transfer between GPU memory and SSDs. %Cognitive SSD~\cite{cog-ssd} offloads requests of unstructured data retrial directly to NVMe devices, which reduces data I/O between SSD, memory and caches. HolisticGNN~\cite{HolisticGNN} accelerates graph neural networks processing by incorporating in-storage processing directly on NVMe devices.
Zero-infinity~\cite{zero-infinity} treats NVMe as the backbone slow memory and transfers data into GPU HBM memory when needed, which empowers giant deep learning model to be trained on limited number of GPUs. Different from above work, \name{} is specially designed for deep learning checkpoiting workload and leveraging data parallel model replicas to further reduce data persistence overhead.
\section{Conclusion}
To the best of our knowledge, \name is the first system effort that utilizes NVMe capability to improve efficiency of DL model checkpointing. \name{} leverages NVMe optimizations and data parallel writes to near-linearly scale checkpointing performance. Furthermore, \name{} leverages DL domain knowledge to overlap model checkpointing with training iteration to reduce checkpointing stalls. Evaluation results show that, \name{} can checkpoint large models frequently, on every training iteration, with negligible overhead. %enables model checkpointing without cost of training efficiency.  

\bibliographystyle{plain}
\bibliography{main}

\clearpage
\section{Appendix}
\begin{figure}[ht]
    \centering
    \subfigure[Single GPU, \textit{32MB Checkpoint}]{\includegraphics[width=0.23\textwidth]{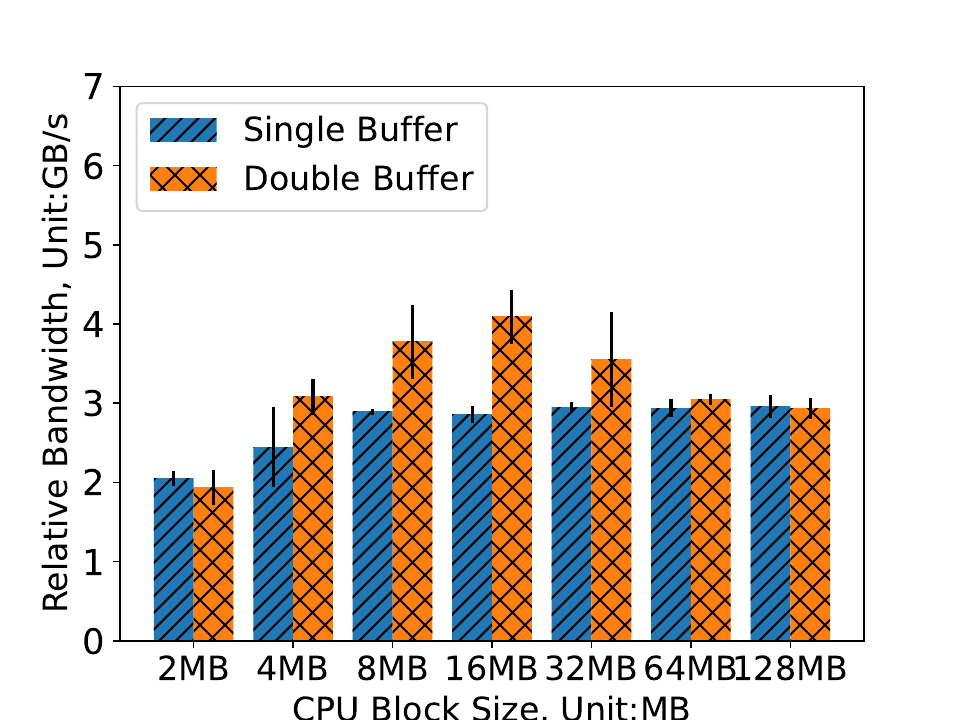}} 
    \subfigure[Single GPU, \textit{64MB Checkpoint}]{\includegraphics[width=0.23\textwidth]{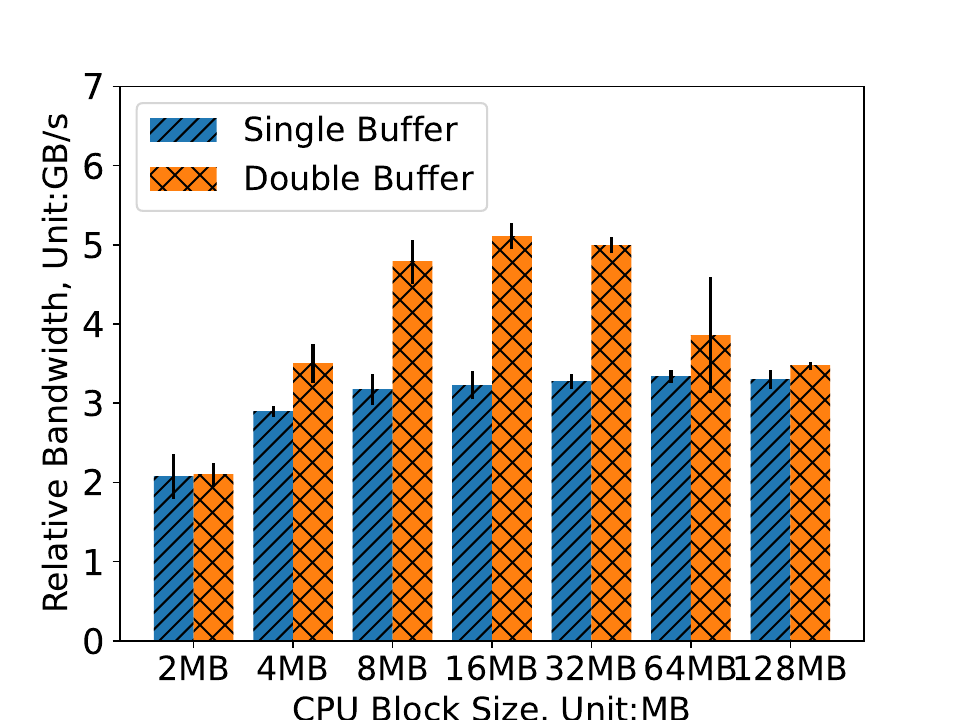}} 
    \caption{Varying {\it IO Buffer} size (32MB and 64MB) on single GPU }
    \label{fig:block_size1}
\end{figure}

\begin{figure}[ht]
    \centering
    \subfigure[Single GPU, \textit{128MB Checkpoint}]{\includegraphics[width=0.23\textwidth]{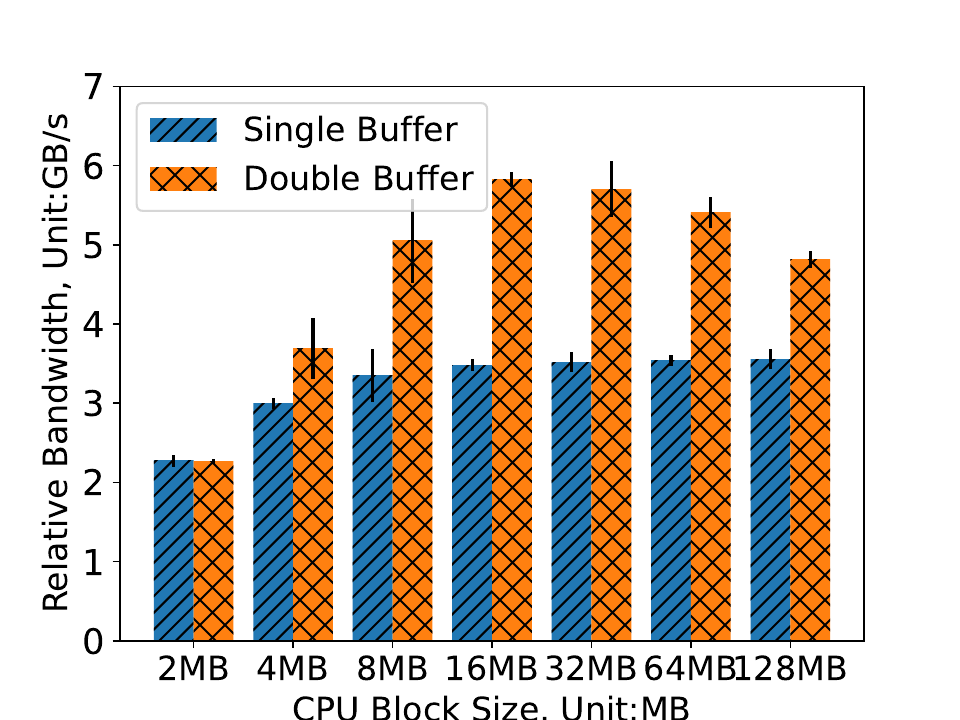}} 
    \subfigure[Single GPU, \textit{256MB Checkpoint}]{\includegraphics[width=0.23\textwidth]{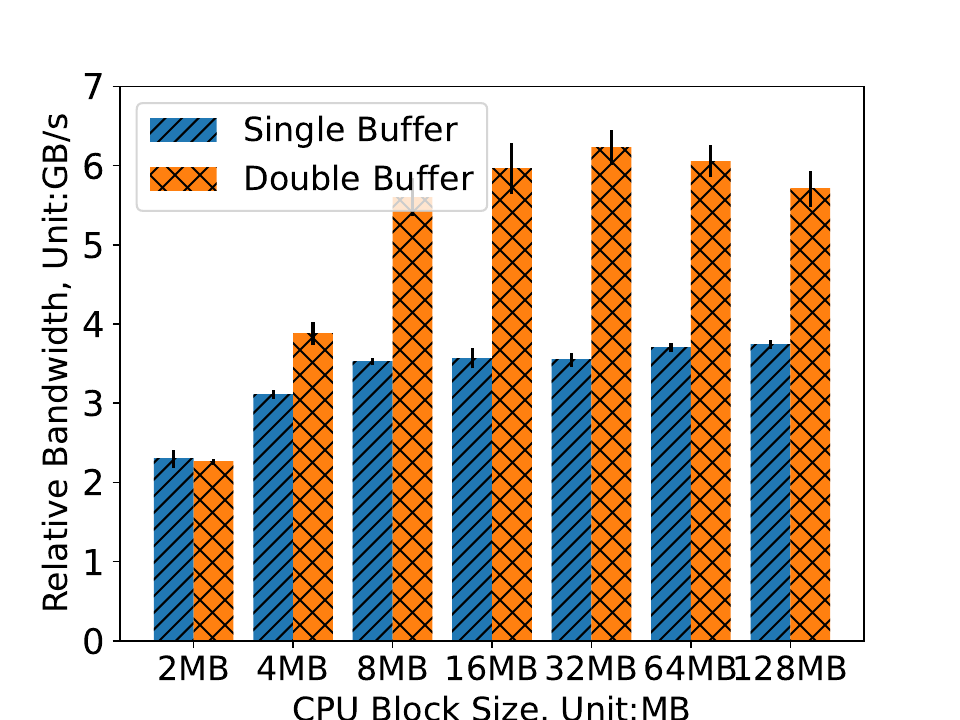}} 
    \caption{Varying {\it IO Buffer} size (128MB and 256MB) on single GPU}
    \label{fig:block_size2}
\end{figure}

\begin{figure}[ht]
    \centering
     \subfigure[\textit{1 Nodes (2 Sockets).}]{\includegraphics[width=0.23\textwidth]{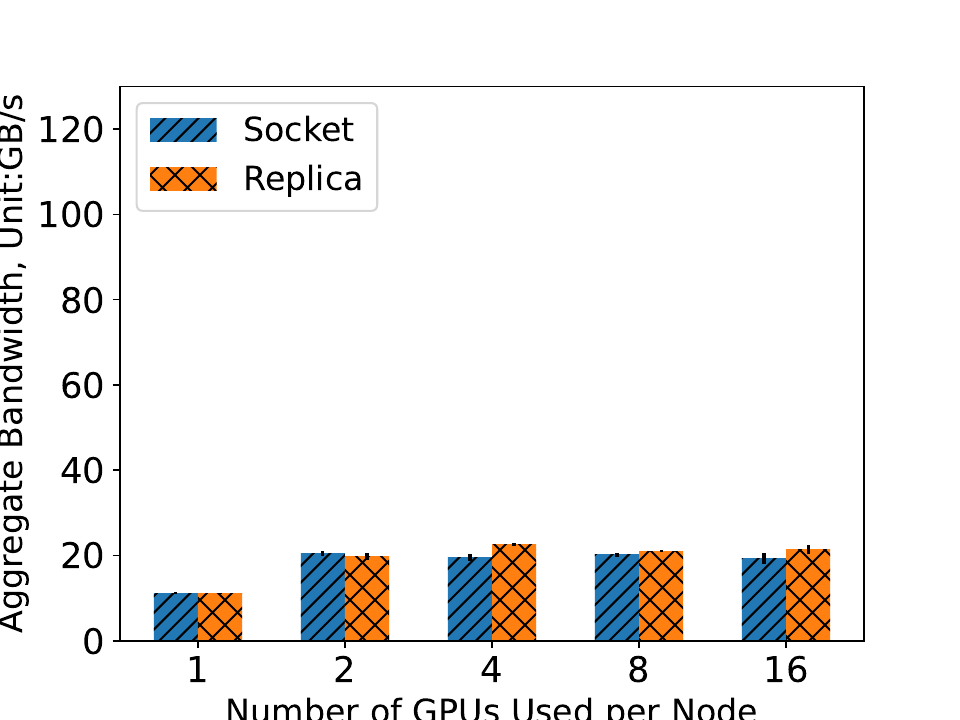}} 
    \subfigure[\textit{4 Nodes (8 Sockets).}]{\includegraphics[width=0.23\textwidth]{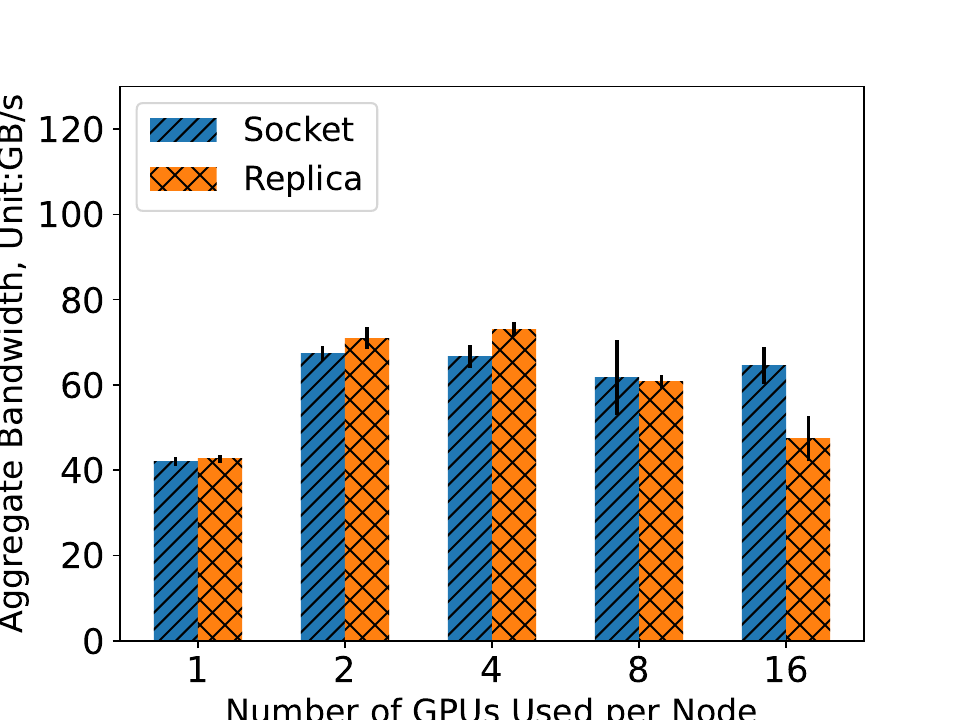}} 

    \caption{Parallel checkpointing of {\it gpt3-0.7b}.}
    \label{fig:multi_nodes1}
\end{figure}

This section presents more results of micro-benchmark (\S\ref{subsec:micro_benchmark}). In particular, we report the IO buffer size in the range of $32$MB--$256$MB on a single GPU in Figures \ref{fig:block_size1} and \ref{fig:block_size2}. Moreover, we also report the multi-node experiment results in Figure \ref{fig:multi_nodes1} for 1 node and 4 nodes. The conclusions can be found in \S\ref{subsec:micro_benchmark}.

%%%%%%%%%%%%%%%%%%%%%%%%%%%%%%%%%%%%%%%%%%%%%%%%%%%%%%%%%%%%%%%%%%%%%%%%%%%%%%%%
\end{document}